\documentclass[12pt]{article} \textwidth 6.25in \textheight 9in
\usepackage{latexsym}
\usepackage{graphics}
\usepackage{graphicx}
\usepackage{latexsym}
\usepackage{amssymb,amsmath}
\usepackage{epsfig}
\usepackage{url}
\makeatletter

\newcommand{\singlespacing}{\let\CS=\@currsize\renewcommand{\baselinestretch}{1}\tiny\CS}
\newcommand{\oneandahalfspacing}{\let\CS=\@currsize\renewcommand{\baselinestretch}{1.25}\tiny\CS}
\newcommand{\doublespacing}{\let\CS=\@currsize\renewcommand{\baselinestretch}{1.5}\tiny\CS}
\newcommand{\beqa}{\begin{eqnarray}}
\newcommand{\eeqa}{\end{eqnarray}} \newcommand{\mb}{\mathbf}
\newcommand{\beqan}{\begin{eqnarray*}}
\newcommand{\eeqan}{\end{eqnarray*}}
\newcommand{\beq}{\begin{equation}} \newcommand{\eeq}{\end{equation}}

\newcommand{\df}{\stackrel{{\rm def}}{=}}
\newcommand{\prob}{{\mathbb {P}}}

\newcommand{\expect}{{\mathbb {E}}}

\newtheorem{lemma}{Lemma}[section]
\newtheorem{defn}{Definition}[section]
\newtheorem{thm}[lemma]{Theorem}
\newtheorem{corollary}{Corollary}[section]
\newtheorem{prop}{Proposition}[section]

\title{Cognitive Radio: An Information-Theoretic Perspective}
\author{Aleksandar~Jovi\v{c}i\'{c}
and Pramod Viswanath
\thanks{A. Jovi\v{c}i\'{c} and P. Viswanath are with the
 department of Electrical and  Computer Engineering at the
 University of Illinois at  Urbana-Champaign.
 Email: {\tt \{jovicic,pramodv\}@uiuc.edu}}
\thanks{This research was supported in part by
the National Science Foundation under grant CCR-0312413
 and a grant from Motorola Inc.\ as part of the Motorola
Center for Communication.}}

\begin{document}

\maketitle
\begin{abstract}
Cognitive radios have been proposed as a means to implement
efficient reuse of the licensed spectrum. The key feature of a
cognitive radio is its ability to recognize the primary (licensed)
user and adapt its communication strategy to minimize the
interference that it generates. We consider a communication scenario
in which the primary and the cognitive user wish to communicate to
different receivers, subject to mutual interference. Modeling the
cognitive radio as a transmitter with side-information about the
primary transmission, we characterize the largest rate at which the
cognitive radio can reliably communicate under the constraint that
(i) {\em no interference} is created for the primary user, and (ii)
the primary encoder-decoder pair is oblivious to the presence of the
cognitive radio.
\end{abstract}

%\begin{keywords}

%\end{keywords}

\section{Introduction}
Observing a severe
under-utilization of the licensed spectrum, the FCC has
recently recommended \cite{FCC1, FCC2} that significantly greater
spectral efficiency could be realized by deploying wireless devices
that can coexist with the incumbent licensed (primary) users,
generating minimal interference while somehow taking advantage of
the available resources. Such devices could, for instance, form
real-time secondary markets \cite{PS03} for the licensed spectrum
holders of a cellular network or even, potentially, allow a complete
secondary system to simultaneously operate in the same frequency
band as the primary. The characteristic feature of these {\em
cognitive radios} would be their ability to {\em recognize} their
communication environment and {\em adapt} the parameters of their
communication scheme to maximize the quality of service for the
secondary users while minimizing the interference to the primary
users.

In this paper, we study the fundamental limits of performance of
wireless networks endowed with cognitive radios. In particular, in
order to understand the ultimate system-wide benefits of the
cognitive nature of such devices, we assume that the cognitive radio
has non-causal knowledge of the codeword of the primary user in its
vicinity\footnote{Note that this does not imply that the cognitive
user can decode the {\em information} that the primary user is
communicating since there are secure encryption protocols running at
the application layer. The decoded codeword is a meaningless stream
of bits for the cognitive user.}; in this, we are motivated by the
model proposed in \cite{Harvard}. We address the following
fundamental question:

{\em What is the largest rate that the cognitive radio can achieve
under the constraint that }
\begin{itemize}
\item[(i)] {\em it
generates no interference for the primary user in its vicinity, and}
\item[(ii)] {\em the primary receiver uses a single-user decoder, just as it would in the
absence of the cognitive radio?}
\end{itemize}

We will refer to these two imperative constraints as the {\em
coexistence conditions} that a cognitive secondary system must
satisfy.
\begin{figure}[h!]
         \centerline{
             \includegraphics[width=10cm]{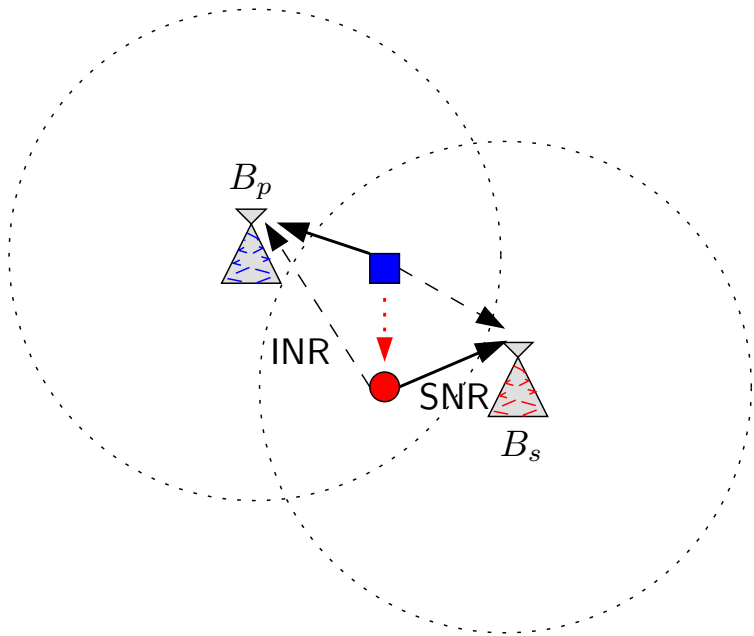}
             } \caption{A possible arrangement of the primary and
           secondary receivers, base-stations $B_p$ and $B_s$,
           respectively. The cognitive secondary user is represented by the
           circle and the primary user is represented by the
           square. The side-information path is depicted by the dotted line.}
\end{figure}\label{xi-network}

Of central interest to us is the communication scenario illustrated
in Fig.~\ref{xi-network}: The primary user wishes to communicate to
the primary base-station $B_p$. In its vicinity is a secondary user
equipped with a cognitive radio that wishes to transmit to the
secondary base-station $B_s$. We assume that the cognitive radio has
obtained the message of the primary user. The received
signal-to-noise ratio of the cognitive radio's transmission at the
secondary base-station is denoted by ${\sf SNR}$. The transmission
of the cognitive radio is also received at $B_p$, and the
signal-to-noise ratio of this interfering signal is denoted by ${\sf
INR}$ (interference-to-noise ratio). If the cognitive user is close
to $B_p$, ${\sf INR}$ could potentially be large.

Our main result is the characterization of the largest rate at which
the cognitive radio can reliably communicate with its receiver $B_s$
under the coexistence conditions and in the
``low-interference-gain'' regime in which ${\sf INR}\leq {\sf SNR}$.
This regime is of practical interest since it models the realistic
scenario in which the cognitive radio is closer to $B_s$ than to
$B_p$. Moreover, we show that the capacity achieving strategy is for
the cognitive radio to perform {\em precoding} for the primary
users's codeword and transmit over the {\em same} time-frequency
slot as that used by the primary radio.

To prove our main result, we allow the primary and secondary systems
to {\em cooperate} and jointly design their encoder-decoder pairs
and then show that the optimal communication scheme for this
cooperative situation has the property that the primary decoder does
not depend on the encoder and decoder used by the secondary system.
This cooperative communication scenario can be thought of as an
interference channel \cite{Carleial}, \cite{Sato}, \cite{Costa2} but
with degraded message sets\footnote{The primary radio has only a
subset of the messages available to the cognitive radio.}:
Achievable schemes for this channel have been first studied in
\cite{Harvard}. A related problem of communicating a single private
message along with a common message to each of the receivers has
been studied in \cite{Maric}.

Furthermore, we exhibit a regime in which joint code design {\em is}
beneficial when one considers the largest set of simultaneously
achievable rates of the primary and cognitive users. We show that,
unlike in the low-interference-gain regime, knowledge of the code
used by the cognitive radio is required by the primary decoder in
order to achieve all the rates in the capacity region of this
interference channel when ${\sf INR}\gg {\sf SNR}$.

The rest of this paper is organized as follows. We first present the
Gaussian {\em cognitive channel} in Section~\ref{section_model}. We
state our main result, the capacity of the cognitive channel in the
low-interference-gain regime ${\sf INR}\leq {\sf SNR}$, in
Section~\ref{sec:main_result}. The proof of our main result is given
in Section~\ref{sec:proof}, where we demonstrate the capacity region
of the underlying interference channel with degraded message sets
which inherently allows for joint code design. We then show that the
benefit of joint code design becomes apparent in the
high-interference-gain regime ${\sf INR}\gg{\sf SNR}$; this is done
in Section~\ref{high_interference_section}. Finally, we study the
system-level implications of the optimal cognitive communication
scheme in Section~\ref{system_implications}.

\section{The Channel Model and Problem Statement}\label{section_model}
\subsection{The cognitive channel}\label{sec:origina_channel}
Consider the following communication scenario which we will refer to
as the {\em cognitive channel}.

\begin{figure}[h!]
\begin{center}
  \includegraphics[width=10cm]{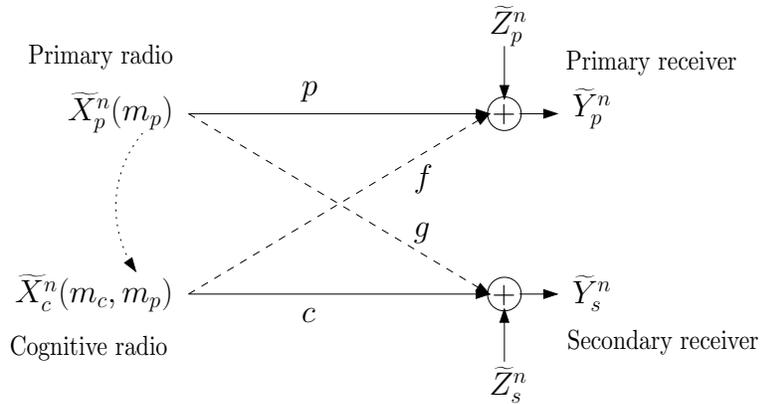}\\
  \caption{The (Gaussian) cognitive channel after $n$ channel uses.
  The dashed lines represent interfering receptions. The dotted line represents
  the side-information path. The power constraints are $\widetilde{P}_p$ and $\widetilde{P}_c$ and noise variances
  are $N_p$ and $N_s$.}\label{IFC}
  \end{center}
\end{figure}

The additive noise at the primary and secondary receivers,
$\widetilde{Z}_p^n:=(\widetilde{Z}_{p,1},\widetilde{Z}_{p,2},\ldots,\widetilde{Z}_{p,n})$
and
$\widetilde{Z}_s^n:=(\widetilde{Z}_{s,1},\widetilde{Z}_{s,2},\ldots,\widetilde{Z}_{s,n})$,
is assumed to be i.i.d. across symbol times $i=1,2,\ldots n$ and
distributed according to $\mathcal{N}(0,N_p)$ and
$\mathcal{N}(0,N_s)$, respectively\footnote{Throughout the paper we
will denote vectors in $\mathbb{R}^n$ by
$X^n:=(X_i,X_2,\ldots,X_n)$}. The correlation between
$\widetilde{Z}^n_p$ and $\widetilde{Z}^n_s$ is irrelevant from the
standpoint of probability of error or capacity calculations since
the base-stations are not allowed to pool their signals. The primary
user has message $m_p\in\{0,1,\ldots,2^{nR_p}\}$ intended for the
primary receiver to decode, the cognitive user has message
$m_c\in\{0,1,\ldots,2^{nR_c}\}$ intended for the secondary receiver
{\em as well} as the message $m_p$ of the primary user. The average
power of the transmitted signals is constrained by $\widetilde{P}_p$
and $\widetilde{P}_c$, respectively: \beqa
\|\widetilde{X}^n_p\|^2\leq n\widetilde{P}_p,\quad
\|\widetilde{X}^n_c\|^2\leq n\widetilde{P}_c. \eeqa

The received signal-to-noise ratios (SNRs) of the desired signals at
the primary and secondary base-station are $p^2\widetilde{P}_p/N_p$
and $c^2\widetilde{P}_c/N_s$, respectively. The received SNRs of the
interfering signals at the primary and secondary base-station (INRs)
are $f^2\widetilde{P}_c/N_p$ and $g^2\widetilde{P}_p/N_s$,
respectively. The constants $(p,c,f,g)$ are assumed to be real,
positive and globally known. The results of this paper easily extend
to the case of complex coefficients (see
Section~\ref{sec:complex_baseband}). The channel can be described by
the pair of per-time-sample equations
 \beqa \widetilde{Y}_p&=&p\widetilde{X}_p+f\widetilde{X}_c+\widetilde{Z}_p,\label{Y_p}\\
\widetilde{Y}_s&=&g\widetilde{X}_p+c\widetilde{X}_c+\widetilde{Z}_s,\label{Y_s}
\eeqa where $\widetilde{Z}_p$ is $\mathcal{N}(0,N_p)$ and
$\widetilde{Z}_s$ is $\mathcal{N}(0,N_s)$.

\subsection{Transformation to standard form}

We can convert every cognitive channel with gains $(p,f,g,c)$, power
constraints $(\widetilde{P}_p,\widetilde{P}_c)$ and noise powers
$(N_p,N_s)$ to a corresponding {\em standard form} cognitive channel
with gains $(1,a,b,1)$, power constraints $(P_p,P_c)$ and noise
powers $(1,1)$, expressed by the pair of equations
\beqa Y_p&=&X_p+aX_c+Z_p,\label{Y_p_standard}\\
Y_s&=&bX_p+X_c+Z_s,\label{Y_s_standard} \eeqa where \beqa
\nonumber a:=\frac{f\sqrt{N_s}}{c\sqrt{N_p}},&~&\quad b:=\frac{g\sqrt{N_p}}{p\sqrt{N_s}},\\
P_p:=\frac{p^2\widetilde{P}_p}{N_p},&~&\quad
P_c:=\frac{c^2\widetilde{P}_c}{N_s}\label{power_transform}.\eeqa The
capacity of this cognitive channel is the same as that of the
original channel since the two channels are related by invertible
transformations\footnote{These transformations were used in
\cite{Carleial}, \cite{Costa} and \cite{Sato}, in the context of the
classical interference channel.} that are given by \beqa &~&X_p:=
\frac{p\widetilde{X}_p}{\sqrt{N_p}},\quad Y_p:=
\frac{\widetilde{Y}_p}{\sqrt{N_p}},\quad
Z_p:= \frac{\widetilde{Z}_p}{\sqrt{N_p}};\\
&~&X_c:= \frac{c\widetilde{X}_c}{\sqrt{N_s}},\quad Y_s:=
\frac{\widetilde{Y}_s}{\sqrt{N_s}},\quad Z_s:=
\frac{\widetilde{Z}_s}{\sqrt{N_s}}. \eeqa
\begin{figure}[h]
\begin{center}
  \includegraphics[width=10cm]{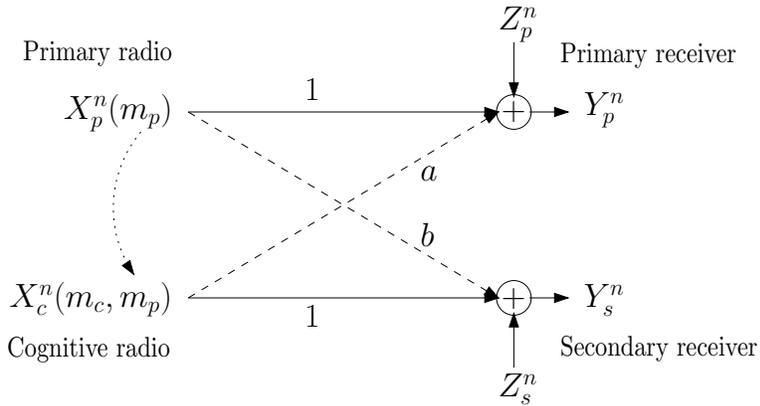}\\
  \caption{The cognitive channel in standard form. The channel gains $(p,f,g,c)$ in the original channel
  are mapped to $(1,a,b,1)$, powers $(\widetilde{P}_p,\widetilde{P}_c)$ are
mapped to $(P_p,P_c)$,
  and noise variances $(N_p,N_s)$ are mapped to $(1,1)$.}\label{IFC_standard}
  \end{center}
\end{figure}

In deriving our main result we will consider this standard form of
the cognitive channel without loss of generality and we will refer
to it as the {\em cognitive $(1,a,b,1)$ channel}.

\subsection{Coding on the cognitive channel}

Let the channel input alphabets of the primary and cognitive radios
be $\mathcal{X}_p=\mathbb{R}$ and $\mathcal{X}_c=\mathbb{R}$,
respectively. Similarly, let the channel output alphabets at the
primary and secondary receivers be $\mathcal{Y}_p=\mathbb{R}$ and
$\mathcal{Y}_s=\mathbb{R}$, respectively.

The primary receiver is assumed to use a standard single-user
decoder to decode $m_p\in\{1,2,\ldots,2^{nR_p}\}$ from $Y^n_p$, just
as it would in the absence of the secondary system: Any decoder
which achieves the AWGN channel capacity, such as the
maximum-likelihood decoder or the joint-typicality decoder, will
suffice. Following standard nomenclature, we say that $R_p$ is {\em
achievable} for the primary user if there exists a sequence (indexed
by $n$) of encoding maps, $E^n_p: \{1,2,\ldots,2^{nR_p}\}\mapsto
\mathcal{X}^n_p$, satisfying $\|X^n_p\|^2\leq nP_p$, and for which
the average probability of decoding error (average over the
messages) vanishes as $n\rightarrow\infty$.

The cognitive radio is assumed to have knowledge of $m_p$, hence we
have the following definition:
\begin{defn}[Cognitive code]\label{defn_cognitive_code}
A cognitive $(2^{nR_c},n)$ code is a choice of an encoding rule
(whose output we denote by $X^n_c$) \beqa
E^n_c&:&\{1,2,\ldots,2^{nR_p}\}\times\{1,2,\ldots,2^{nR_c}\}\rightarrow
\mathcal{X}_c^n,\eeqa such that $\|X^n_c\|^2\leq nP_c$, and a choice
of a decoding rule \beqa D^n_c &:&\mathcal{Y}_s^n\rightarrow
\{1,2,\ldots,2^{nR_c}\}.\eeqa
\end{defn}

The following key definition formalizes the important notion of {\em
coexistence conditions} that the cognitive secondary system must
satisfy.
\begin{defn}[Achievability: cognitive user]\label{defn_cog_achieve}
A rate $R_c$ is said to be achievable for the cognitive user on a
cognitive $(1,a,b,1)$ channel if there exists a sequence of
cognitive $(2^{nR_c},n)$ codes such that the following two
constraints are satisfied:
\begin{enumerate}
\item The average probability of error vanishes as $n\rightarrow
\infty$, i.e., \beqa P^{(n)}_{e,c}\df
\frac{1}{2^{n(R_c+R_p)}}\sum_{i=1,j=1}^n \prob(D^n_c(Y_s^n)\neq
j|m_p=i,m_c=j)\rightarrow 0; \eeqa
\item A rate of $R^*_p\df\frac{1}{2}\log(1+P_p)$ is achievable for the primary user.
\end{enumerate}
\end{defn}
\begin{defn}[Capacity] The {\em capacity} of the cognitive channel is defined to be
the largest achievable rate $R_c$ for the cognitive user.
\end{defn}
Our main result, presented in the following section, precisely
quantifies the capacity of the cognitive channel in the
``low-interference-gain'' regime.
\section{The Main Result}\label{sec:main_result}
If the received SNR of the cognitive radio transmission is lesser at
the primary receiver than at the secondary receiver, we say that the
primary system is affected by a {\em low interference gain}. This is
the case that is most likely to occur in practice since the
cognitive radio is typically closer to its intended receiver (the
secondary base-station) than to the primary base-station. In terms
of the parameters of our problem, this situation corresponds to
$f\sqrt{N_s}\leq c\sqrt{N_p}$ in our original cognitive channel, or,
equivalently, to $a\leq 1$ in the corresponding standard-form
cognitive $(1,a,b,1)$ channel. Our main result is an explicit
expression for the capacity of the cognitive channel in this regime.

\begin{thm}\label{thm:main}
The capacity of the cognitive $(1,a,b,1)$ channel is \beqa
R^*_c=\frac{1}{2}\log(1+(1-\alpha^*)P_c), \eeqa as long as $a\leq
1$. The constant $\alpha^*\in [0,1]$ is defined in
$(\ref{alpha_opt})$.
\end{thm}
Note that Theorem~\ref{thm:main} holds for any $b\in\mathbb{R}$ (or
equivalently any $p,g\in\mathbb{R}$ in the original cognitive
channel).

\section{Proof of the Main Result}\label{sec:proof}
\subsection{The forward part}\label{proof_main_forward}
To show the existence of a capacity-achieving cognitive
$(2^{nR_c^*},n)$ code, we generate a sequence of random codes such
that the average probability of error (averaged over the ensemble of
codes and messages) vanishes as $n\rightarrow \infty$. In
particular, we have the following codes:
\begin{itemize}
\item {\em $E^n_p$ ensemble:} Given
$m_p\in\{1,2,\ldots,2^{nR_p}\}$, generate the codeword
$X^n_p\in\mathbb{R}^n$ by drawing its coordinates i.i.d. according
to $\mathcal{N}(0,P_p)$.
\item {\em $E^n_c$ ensemble:} Since the cognitive radio knows $m_p$ as
well as $E^n_p$, it can form $X^n_p$ and perform superposition
coding as follows: \beqa X^n_c=\hat{X}^n_c+\sqrt{\frac{\alpha
P_c}{P_p}}X^n_p,\label{superposition} \eeqa where $\alpha\in [0,1]$.
The codeword $\hat{X}^n_c$ encodes $m_c\in\{1,2,\ldots,2^{nR_c}\}$
and is generated by performing {\em Costa precoding} \cite{Costa}
(also known as {\em dirty-paper coding}) treating
$(b+\sqrt{\alpha\frac{P_c}{P_p}})X^n_p$ as non-causally known
interference that will affect the secondary receiver in the presence
of $\mathcal{N}(0,1)$ noise. The encoding is done by {\em random
binning} \cite{Costa}.
\item $D^n_c$: Costa decoder (having knowledge of the binning encoder $E^n_c$) \cite{Costa}.
\end{itemize}

The key result of Costa \cite{Costa2} is that, using the dirty-paper
coding technique, the maximum achievable rate is the same as if the
interference was also known at the receiver, i.e., as if it were
absent altogether. The characteristic feature of this scheme is that
the resulting codeword $\hat{X}_c^n$ is statistically independent of
$X^n_p$ and is i.i.d. Gaussian. To satisfy the average power
constraint of $P_c$ on the components of $X^n_c$, each coordinate of
$\hat{X}^n_c$ must, in fact, be $\mathcal{N}(0,(1-\alpha)P_c)$.
Hence, the primary receiver can treat $\hat{X}^n_c$ as independent
Gaussian noise. Using standard methodology, it can be shown that the
average probability of error for decoding $m_p$ (averaged over the
code ensembles and messages) vanishes, as $n\rightarrow \infty$, for
all rates $R_p$ below \beqa
\frac{1}{2}\log\left(1+\frac{(\sqrt{P_p}+a\sqrt{\alpha
P_c})^2}{1+a^2(1-\alpha)P_c}\right). \eeqa Similarly, the average
probability of error in decoding $m_c$ vanishes for all rates $R_c$
below \beqa \frac{1}{2}\log(1+(1-\alpha)P_c). \eeqa However, in
order to ensure that a given rate is {\em achievable} for the
cognitive user in the sense of Definition~\ref{defn_cog_achieve}, we
must have that \beqa
\frac{1}{2}\log\left(1+\frac{(\sqrt{P_p}+a\sqrt{\alpha
P_c})^2}{1+a^2(1-\alpha)P_c}\right)=\frac{1}{2}\log(1+P_p)=:R_p^*.
\label{no_interference}\eeqa Observe that, if $a=0$, any choice of
$\alpha\in [0,1]$ will satisfy $(\ref{no_interference})$: in this
case we should set $\alpha^*=0$ to maximize the rate achievable for
the cognitive user. For $0< a \leq 1$, by the Intermediate Value
Theorem, this quadratic equation in $\alpha$ always has a unique
root in $[0,1]$: \beqa
\alpha^*=\left(\frac{\sqrt{P_p}\left(\sqrt{1+a^2P_c(1+P_p)}-1\right)}
{a\sqrt{P_c}(1+P_p)}\right)^\frac{1}{2}.\label{alpha_opt} \eeqa

Finally, since the code-ensemble-averaged (and message-averaged)
probabilities of error vanish, there must exist a particular
sequence of cognitive codes and primary encoders for which the
(message-averaged) probabilities of error vanish as well. Hence,
$R^*_c=\frac{1}{2}\log(1+(1-\alpha^*)P_c)$ is achievable for the
cognitive user in the sense of Definition~\ref{defn_cog_achieve}.

\subsection{The converse part}
\subsubsection{Proof outline}
In order to prove the converse to our main result we will first
relax the constraints of our problem and allow for {\em joint}
primary and cognitive code design. This relaxation leads naturally
to an interference channel with degraded message sets\footnote{The
primary user knows $m_p$ while the cognitive user knows
$\{m_p,m_c\}$, hence the primary user has a subset of the messages
available to the cognitive user.}, which we will abbreviate as
IC-DMS for convenience.

Our approach is to first characterize the capacity region of the
IC-DMS, i.e., the largest set of rate tuples $(R_p,R_c)$ at which
joint reliable communication can take place. We then make the key
observation that the joint coding scheme that achieves all the rate
tuples in the capacity region of the IC-DMS has the property that
the decoder at the primary receiver is a standard single-user
decoder. Furthermore, we show that there exists a point
$(R_p,R_c)=(R_p^*,R_c^*)$ on the boundary of the capacity region of
the IC-DMS, where $R_p^*=\frac{1}{2}\log(1+P_p)$ and
$R^*_c=\frac{1}{2}\log(1+(1-\alpha^*)P_c)$ with $\alpha^*$ given by
$(\ref{alpha_opt})$. We then conclude that $R_c=R_c^*$ is the
capacity of the corresponding cognitive channel.

\subsubsection{Joint code design: The IC-DMS}

The input-output equations of the IC-DMS, as for the cognitive
channel, are given by $(\ref{Y_p})$, $(\ref{Y_s})$ with the standard
form given by $(\ref{Y_p_standard})$, $(\ref{Y_s_standard})$. We
will denote the IC-DMS in standard form by ``$(1,a,b,1)$-IC-DMS''.

\begin{defn}[IC-DMS code]\label{ICDMS_code}
A $(2^{nR_p},2^{nR_c},n)$ code for the $(1,a,b,1)$-IC-DMS
is a choice of an encoding rule and a decoding rule: The encoding
rule is a pair of maps (whose outputs we denote by $X^n_p$ and
$X^n_c$, respectively)
\beqa e_p^n&:&\{1,2,\ldots,2^{nR_p}\}\rightarrow \mathcal{X}_p^n,\label{encoder_p}\\
e_c^n&:&
\{1,2,\ldots,2^{nR_p}\}\times\{1,2,\ldots,2^{nR_c}\}\rightarrow
\mathcal{X}_c^n, \label{encoder_c}\eeqa such that $\|X^n_p\|^2\leq
nP_p$ and $\|X^n_c\|^2\leq nP_c$. The decoding rule is a pair of
maps \beqa
d^n_p&:&\mathcal{Y}_p^n\rightarrow \{1,2,\ldots,2^{nR_p}\},\label{decoder_p}\\
d^n_c&:&\mathcal{Y}_s^n\rightarrow
\{1,2,\ldots,2^{nR_c}\}.\label{decoder_c} \eeqa \end{defn} Given
that the messages selected are $(m_p=i,m_c=j)$, an error occurs if
$d^n_p(Y^n_p)\neq i$ or $d^n_c(Y^n_s)\neq j$.

\begin{defn}[Achievability: IC-DMS]\label{ICDMS_achievability} A rate vector $(R_p, R_c)$ is said to
be {\em achievable} if there exists a sequence of
$(2^{nR_p},2^{nR_c},n)$ codes such that the average probability of
error at each of the receivers vanishes as $n\rightarrow \infty$,
i.e., \beqa \widetilde{P}^{(n)}_{e,p}&\df&
\frac{1}{2^{n(R_c+R_p)}}\sum_{i=1,j=1}^n
\prob(d^n_p(Y_p^n)\neq i|m_p=i,m_c=j)\rightarrow 0,\\
P^{(n)}_{e,s}&\df& \frac{1}{2^{n(R_c+R_p)}}\sum_{i=1,j=1}^n
\prob(d^n_c(Y_s^n)\neq j|m_p=i,m_c=j)\rightarrow 0.\eeqa\end{defn}
\begin{defn}[Capacity region] The capacity region of the IC-DMS is
the closure of the set of achievable rate vectors $(R_p,R_c)$.
\end{defn}
\subsubsection{The capacity region of the IC-DMS under a low interference gain}
The following theorem characterizes the capacity region of the
$(1,a,b,1)$-IC-DMS with $a\leq 1$ and arbitrary $b\in\mathbb{R}$.
\begin{thm}\label{mu_less_one}
The capacity region of the $(1,a,b,1)$-IC-DMS with $a\leq 1$ and
$b\in \mathbb{R}$ is given by the union, over all $\alpha\in [0,1]$,
of the rate regions \beqa 0\leq
R_p&\leq&\frac{1}{2}\log\left(1+\frac{(\sqrt{P_p}+a\sqrt{\alpha
P_c})^2}{1+a^2(1-\alpha)P_c}\right),\label{R_p_low}\\
0\leq
R_c&\leq&\frac{1}{2}\log\left(1+(1-\alpha)P_c\right).\label{R_c_low}
\eeqa
\end{thm}

\vspace{0.1in} \noindent{\em Proof of achievability}: The random
coding scheme described in the forward part of the proof of
Theorem~\ref{thm:main} (Section~\ref{proof_main_forward}) achieves
the rates $(\ref{R_p_low})$ and $(\ref{R_c_low})$ stated in the
theorem. We emphasize that, in this scheme, the primary receiver
employs a single-user decoder.

\noindent {\em Proof of converse:} See
Appendix~\ref{sec:proof_converse_main_result}.

\subsubsection{The capacity of the cognitive channel under a low interference gain}
The proof of Theorem~\ref{mu_less_one} reveals that the jointly
designed code that achieves all the points on the boundary of the
capacity region of the IC-DMS is such that the primary receiver uses
a standard single-user decoder, just as it would {\em in the
absence} of the cognitive radio. In other words, the primary decoder
$d^n_p$ does not depend on $e^n_c$ and $d^n_c$. Thus, in order to
find the largest rate that is achievable by the cognitive user in
the sense of Definition~\ref{defn_cog_achieve} we can without loss
of generality restrict our search to the boundary of the capacity
region of the underlying IC-DMS. Hence, to find this capacity of the
cognitive channel, we must solve for the positive root of the
quadratic equation $(\ref{no_interference})$ in $\alpha$. The
solution is given by $\alpha^*$ in $(\ref{alpha_opt})$, hence the
capacity is \beqa R_c^*=\frac{1}{2}\log(1+(1-\alpha^*)P_c). \eeqa
Thus we have established the proof of Theorem~\ref{thm:main}.\hfill
$\square$

The proof of the converse of Theorem~\ref{mu_less_one} allows us to
characterize the sum-capacity of the $(1,a,b,1)$-IC-DMS for any
$a\geq 1$ and the entire capacity region if $a$ is sufficiently
large. These two ancillary results are shown in the following
section.
\subsubsection{The high-interference-gain regime}
{\bf The sum-capacity for $a\geq 1$}
\begin{corollary}\label{cor_sum}
The maximum of $R_p+R_c$ over all $(R_p,R_c)$ in the capacity region
of the $(1,a,b,1)$-IC-DMS with $a\geq 1$ and $b\in \mathbb{R}$ is
achieved with $\alpha=1$ in $(\ref{R_p_low})$ and $(\ref{R_c_low})$,
i.e., \beqa
C_{\text{sum}}(a)=\frac{1}{2}\log\left(1+\left(\sqrt{P_p}+a\sqrt{P_c}\right)^2\right).\label{C_sum}
\eeqa
\end{corollary}
\noindent {\em Proof:} See Appendix~\ref{proof_sum_capacity}

Contrary to the development so far, in the following section we will
observe that, in the very-high-interference-gain regime, the optimal
(jointly designed) IC-DMS code is such that the primary decoder
$d^n_p$ depends on the cognitive encoder $e_c^n$.

\noindent{\bf The benefit of joint code
design}\label{high_interference_section}

\noindent When the interference gain at the primary receiver due to
the cognitive radio transmissions (parameter $a$) is sufficiently
large, the optimal decoder at the primary receiver of the IC-DMS is
one that decodes the message of the cognitive user before decoding
the message of the primary user.

First, we demonstrate an achievable scheme in the following lemma.

\begin{lemma}\label{lemma_achievable_high}
Consider the cognitive $(1,a,b,1)$-interference channel. For every
$\alpha\in [0,1]$, the rate pair $(R_p,R_c)$ satisfying \beqa
R_p=\hat{R}_p(\alpha)&\df&\frac{1}{2}\log\left(1+\left(\sqrt{P_p}+a\sqrt{\alpha
P_c}\right)^2\right),\label{R_p_high_achievable}\\
R_c=\hat{R}_c(\alpha)&\df&\frac{1}{2}\log\left(1+\frac{(1-\alpha)P_c}{1+(b\sqrt{P_p}+\sqrt{\alpha
P_c})^2}\right), \label{R_c_high_achievable}\eeqa is achievable as
long as \beqa a\geq \frac{\sqrt{\alpha P_p P_c}}{K(\alpha)}+
\sqrt{K(\alpha)+P_p\left(1+(b\sqrt{P_p}+\sqrt{\alpha
P_c})^2\right)},\label{a_constraint} \eeqa where $K(\alpha)\df
1+b^2P_p+2b\sqrt{\alpha P_p P_c}$.
\end{lemma}

\vspace{0.1in} \noindent{\em Proof:} The primary transmitter forms
$X^n_p$ by drawing its coordinates i.i.d. according to
$\mathcal{N}(0,P_p)$. Since the cognitive radio knows $m_p$ and
$e^n_p$ it forms $X^n_p$ then generates $X^n_c$ by superposition
coding: \beqan X^n_c=\hat{X}^n_c+\sqrt{\frac{\alpha P_c}{P_p}}X^n_p,
\eeqan where $\hat{X}^n_c$ is formed by drawing its coordinates
i.i.d. according to $\mathcal{N}(0,\sqrt{(1-\alpha)P_c})$ for some
$\alpha\in [0,1]$. The decoder $d^n_p$ at the primary receiver first
decodes $m_c$ treating $(1+a\sqrt{\alpha P_c/P_p})X^n_p$ as
independent Gaussian noise. It then reconstructs $a\hat{X}^n_c$
(which it can do because it knows $e^n_c$) and subtracts off its
contribution from $Y^n_p$ before decoding $m_p$. The decoding rule
$d^n_c$ at the secondary receiver is simply to decode $m_c$ treating
$(b+\sqrt{\alpha P_c/P_p})X^n_p$ as independent Gaussian noise. The
rates achievable with this scheme are then exactly given by
$(\ref{R_p_high_achievable})$ and $(\ref{R_c_high_achievable})$,
provided that the rate at which the primary receiver can decode the
cognitive user's message is not the limiting factor, i.e., \beqan
\frac{(1-\alpha)P_c}{1+(b\sqrt{P_p}+\sqrt{\alpha P_c})^2} \leq
\frac{a^2(1-\alpha)P_c}{1+\left(\sqrt{P_p}+a\sqrt{\alpha
P_c}\right)^2}.\eeqan Solving this quadratic inequality for $a$, we
find that the condition is satisfied only when $a$ satisfies
inequality $(\ref{a_constraint})$ stated in the theorem. \hfill
$\square$

\begin{thm}\label{thm:high_interference}
A point $(R_p,R_c)$ is on the boundary of the capacity region of the
cognitive $(1,a,b,1)$-interference channel if there exists
$\alpha\in [0,1]$ such that
\begin{enumerate}
\item $(R_p,R_c)=(\hat{R}_p(\alpha),\hat{R}_c(\alpha))$ where $\hat{R}_p(\alpha)$
and $\hat{R}_c(\alpha)$ are defined in $(\ref{R_p_high_achievable})$
and $(\ref{R_c_high_achievable})$, respectively,
\item $a$ and $b$ satisfy the condition given in $(\ref{a_constraint})$, and
\item $b\leq b_{\max }(\mu_\alpha,a)$ where
$\mu_\alpha\df
-\left.\frac{d^-\hat{R}_c(x)}{d\hat{R}_p(x)}\right|_{x=\alpha}$ and
$b_{\max}(\mu,a)$ is defined in
Appendix~\ref{sec:converse_high_interference}.
\end{enumerate}
\end{thm}
\noindent{\em Proof of achievability:} Given in
Lemma~\ref{lemma_achievable_high}.

\noindent{\em Proof of converse:} Given in
Appendix~\ref{sec:converse_high_interference}.

Observe that Theorem~\ref{thm:high_interference} characterizes the
entire capacity region of the $(1,a,b,1)$-IC-DMS with $a\geq
\sqrt{P_p P_c}/K(1)+ \sqrt{K(1)+P_p\left(1+(b\sqrt{P_p}+\sqrt{
P_c})^2\right)}$ and $b\leq b_{\max}(\mu_\alpha,a)$.

\section{System-level Considerations}\label{system_implications}
In this section we use our results on the capacity-achieving
cognitive communication scheme to derive insight into a practical
implementation of cognitive radios.

\subsection{Properties of the optimal scheme}
\subsubsection{Avoiding the ``hidden-terminal'' problem} The network
of Fig.~\ref{xi-network} models the situation in which the
geographic location of $B_s$ is not assigned in accordance with any
centralized cell-planning policy and it can be arbitrarily close to
$B_p$. Consequently, the secondary users that are in close proximity
to $B_p$ could potentially cause significant interference for the
primary system if the secondary system is to operate over the same
frequency band.

One possible adaptive communication scheme that the cognitive radio
could employ in order to avoid interfering with the primary user in
its vicinity would be to restrict its transmissions to only the
time-frequency slots which are not occupied by the signals of the
detected primary radio. Indeed, this idea of ``opportunistic''
orthogonal communication was what led to the birth of the notion of
cognitive radio. However, one drawback of such a protocol is that
the cognitive radio would very likely cause interference to other,
more distant, primary users whose presence -- i.e., time-frequency
locations -- it could not detect. The degradation in overall
performance of the primary system due to this ``hidden-terminal''
problem could potentially be significant\footnote{Classical RTS/CTS
solutions to this problem are not viable since they require that the
primary system {\em ask} for access to the very spectrum that it
owns.}, especially in the context of OFDMA \cite{Wimax}, \cite{ITU}
where the primary users are allocated orthogonal time-frequency
slots and the SINR required for decoding is typically large.

Contrary to this, we find that the optimal strategy is for the
cognitive radio to simultaneously transmit in the same frequency
slot as that used by the primary user in its vicinity. An immediate
benefit of this scheme is that, if the transmissions of different
primary users are mutually orthogonal, the cognitive radio can {\em
only} (potentially) affect the performance achievable by the primary
radio whose codeword it has decoded. Furthermore, we know that a
proper tuning of the parameter $\alpha$ can, in fact, ensure that
the primary user's rate is unaffected.

\subsubsection{Robustness to noise statistics} All our results have
been derived under the assumption that the noise affecting the
receivers, $Z^n_p$ and $Z^n_s$, is i.i.d. Gaussian. In
\cite{Lapidoth2} it was shown that using a Costa encoder-decoder
pair that is designed for additive i.i.d Gaussian noise on a channel
with arbitrary (additive) noise statistics will cause no loss in the
achievable rates.\footnote{Note that this is an achievability
result: The capacity of the channel with this arbitrary noise could
be larger but a different code would be required to achieve it.}
Combined with the similar classical result for the standard AWGN
channel \cite{Lapidoth1}, we see that the maximal rate expressed in
Theorem~\ref{thm:main} is achievable for all noise distributions.

\subsection{Obtaining the side-information}
In practice, the cognitive radio must obtain the primary radio's
codeword in a causal fashion -- its acquisition thus introducing
delays in the cognitive radio transmissions\footnote{Under a
half-duplex constraint the cognitive radio must first ``listen'' in
order to decode the primary message before it can use this
side-information for its own transmission.}. In a typical situation,
due to its relative proximity to the primary user, the cognitive
radio can receive the primary transmissions with a greater received
SNR than that experienced by the primary receiver. Hence, it seems
plausible that the cognitive radio could decode\footnote{The
cognitive radio is assumed to know the encoder of the primary user.}
the message of the primary user in fewer channel uses than are
required by the primary receiver. Recent work in distributed
space-time code design \cite{MOT05} indicates that this overhead
decoding delay is negligible if the cognitive radio has as little as
a $10$ dB advantage in the received SNR over the primary receiver.

\subsection{Extension to complex
baseband}\label{sec:complex_baseband} The results of this paper can
easily be extended to the case in which the channel gains are
complex quantities, i.e., $p,f,g,c\in\mathbb{C}$ in the case of the
original cognitive $(p,f,g,c)$ channel with power constraints
$(P_p,P_c)$ and noise variances $(N_p,N_s)$, as defined in
Section~\ref{sec:origina_channel}. However, the optimal cognitive
encoder rule $(\ref{superposition})$ must change slightly: The
superposition scheme takes the form \beqa
X^n_c=\hat{X}_c^n+\frac{f^*}{|f|}e^{j\theta_p}\sqrt{\alpha\frac{P_c}{P_p}}X^n_p,
\label{complex_superposition} \eeqa where $p=|p|e^{j\theta_p}$. The
codeword $\hat{X}^n_c$ is again generated by Costa precoding, but
the assumed interference at the secondary receiver is now \beqa
\left(\frac{g}{c}+
\frac{f^*}{|f|}e^{j\theta_p}\sqrt{\alpha\frac{P_c}
{P_p}}\right)X^n_p,\label{known_interference}\eeqa and the assumed
noise is $\mathcal{CN}(0,N_s/|c|^2)$. The factor $e^{j\theta_p}$ in
$(\ref{complex_superposition})$ essentially implements transmit
beamforming to the primary receiver, hence ensuring that all the
rates given by \beqa 0\leq
R_p&\leq&\log\left(1+\frac{\left(|p|\sqrt{P_p}+|f|\sqrt{\alpha
P_c}\right)^2}{N_p+|f|^2(1-\alpha)P_c}\right),\\
0\leq R_c&\leq&\log\left(1+\frac{|c|^2(1-\alpha)P_c}{N_s}\right),
\eeqa are achieved in the underlying IC-DMS. As before, we can then
choose $\alpha=\alpha^*$ (determined by $(\ref{alpha_opt})$), so
that $R_c^*=\log(1+|c|^2(1-\alpha^*)P_c/N_s)$ is achievable in the
spirit of Definition~\ref{defn_cog_achieve} but with
$R_p^*=\log(1+|p|^2P_p/N_p)$.

\subsection{Communicating without channel-state feedback from the primary base-station}
In order to perform the complex base-band superposition coding
scheme $(\ref{complex_superposition})$ and, implicitly, the Costa
precoding for known interference $(\ref{known_interference})$, the
cognitive radio must know each of the four parameters $g$, $c$, $f$
and $p$, both in magnitude and phase. %While the magnitudes of these
%parameters will typically be changing very slowly over the
%time-scale of communication, the phase could change very rapidly in
%response to even the slightest change in the relative position of
%the transmitters and the receivers.
To obtain estimates for $p$ and $f$, the cognitive radio would
require feedback from the primary base-station. In section
Section~\ref{CSI_feedback}, we discuss ways in which the estimation
and feedback of these parameters could be implemented. In this
section, however, we present an alternative (suboptimal) scheme
which requires no feedback from the primary base-station.

Suppose that, after having decoded $X^n_p$, the cognitive radio
transmits the following $n$-symbol codeword: \beqa
X^n_c=\hat{X}^n_c+\sqrt{\alpha\frac{P_c}{P_p}}X^n_p, \eeqa where the
codeword $\hat{X}^n_c$ is generated by Costa precoding for the
interference\beqa
\left(\frac{g}{c}+\sqrt{\alpha\frac{P_c}{P_p}}\right)X^n_p, \eeqa
assuming the presence of $\mathcal{CN}(0,N_s/|c|^2)$ noise at the
secondary base-station.

\begin{itemize}
\item {\em Obtaining $c$:} The parameter $c$ could be estimated at the secondary
base-station by using the cognitive radio's pilot signal or in a
decision-directed fashion. The estimate could then be fed back to
the cognitive radio.

\item {\em Obtaining $g$:} If the secondary base-station
synchronizes to the primary radio's pilot signal, it could estimate
$g$ during the time the cognitive radio is in its silent
``listening'' phase and then feed this estimate back to the
cognitive radio. Alternatively, if the cognitive radio reveals to
the secondary base-station the code used by the primary radio, the
secondary base-station could use the silent ``listening'' phase to
decode a few symbols transmitted by the primary radio thereby
estimating the parameter $g$.
\end{itemize}

 We can express the received discrete-time
base-band signal at the primary base-station at time sample $m$ as
\beqa Y_p[m]&=&p X_p[m]+ f\sqrt{\alpha\frac{P_c}{P_p}}X_p[m-l_c]+
Z_{\text{total}}[m], \eeqa where
$Z_{\text{total}}[m]=f\hat{X}_c[m-l_c]+Z_p[m]$ is the aggregate
noise. The integer $l_c$ accounts for the delay incurred while the
cognitive radio ``listens'' and decodes the primary codeword before
it transmits its own signal. This equation essentially describes a
time-invariant two-tap ISI channel for the primary transmission,
hence we can apply a Rake receiver (in the case the primary system
uses direct-sequence spread-spectrum) or transmit-receive
architectures such as OFDM\footnote{The primary base-station would
most likely already employ one of these schemes as a means of
dealing with the multi-path point-to-point channel between the
primary radio and itself. In the context of OFDM, however, the
cyclic prefix would have to be long enough to account for the extra
delay-spread introduced by the cognitive radio's transmission.} to
extract both a diversity gain of two and a power gain of
$|p|^2\widetilde{P}_p+|f|^2\alpha P_c$ at the primary base-station
(see, for instance, Chapter~3 of \cite{TV_book}, and references
therein). Given $\alpha\in [0,1]$, the rates achievable by the
primary and cognitive users using such a scheme are given by \beqa
0\leq R_p&\leq&\log\left(1+\frac{|p|^2P_p+|f|^2\alpha
P_c}{N_p+|f|^2(1-\alpha)P_c}\right),\\
0\leq R_c&\leq&\log\left(1+\frac{|c|^2(1-\alpha)P_c}{N_s}\right).
\eeqa In order to avoid causing interference to the primary user,
the following equation must be satisfied: \beqa \frac{|p|^2
P_p+|f|^2\alpha
P_c}{N_p+|f|^2(1-\alpha)P_c}=\frac{|p|^2P_p}{N_p},\label{no_interference_diversity}
\eeqa If the cognitive radio tunes its parameter $\alpha$ such that
\beqa \alpha=\frac{|p|^2 P_p/N_p}{1+|p|^2 P_p/N_p},
\label{alpha_diversity}\eeqa this condition will be satisfied, hence
$R_p=R_p^*$. Expression $(\ref{alpha_diversity})$ confirms the
intuitive notion that, if the primary system is operating at high
SNR, the cognitive radio should not interfere with it, i.e.,
$\alpha$ should be close to one.

From $(\ref{alpha_diversity})$, we see that, in order to design the
optimal $\alpha$, the cognitive radio only needs to know the
received SNR of the primary transmission at the primary
base-station: $|p|^2 P_p/N_p$. If the primary system uses a good
(capacity-achieving) AWGN channel code and the cognitive radio knows
this, the cognitive radio can easily compute an estimate of this
received SNR since it knows the rate at which the primary user is
communicating, $R_p$: This estimate is simply given by $e^{R_p}-1$.
Thus, an immediate benefit of this scheme is that the primary
base-station need not feed-back the parameters $f$ and $p$ at all:
The cognitive radio can perform completely autonomously.

Though expression $(\ref{alpha_diversity})$ does not depend on
$|f|$, we can see that $(\ref{no_interference_diversity})$ can
approximately be satisfied even with $\alpha=0$ when $|f|^2$ is very
small. Since the cognitive radio has no information about $|f|$ and,
in practice, may not even be able to obtain $|p|^2P_p/N_p$ (if the
primary system is not using a good AWGN code), a natural way for the
cognitive radio to enter the spectrum of the primary would be by
slowly {\em ramping} up its power $P_c$ from $0$ and decreasing
$\alpha$ from $1$ while simultaneously listening for the Automatic
Repeat Request (ARQ) control signal from the primary base-station.
Once this signal is detected, the cognitive radio would either
slightly decrease $P_c$ or increase $\alpha$ until the primary
base-station stops transmitting ARQs\footnote{This scheme is
analogous to the power control mechanism used in CDMA systems.}.

\subsection{Obtaining the channel-state
information}\label{CSI_feedback} In order to implement the optimal
communication scheme of Costa coding and beamforming
$(\ref{complex_superposition})$, the cognitive radio must obtain
estimates of $p$ and $f$ from the primary base-station. We present
the following simple algorithm for estimation and feedback of these
parameters:
\begin{enumerate}
\item At first, the cognitive user is silent and the primary base-station
broadcasts the current estimate of $p$, call it $\hat{p}$, along
with the primary user's ID on the control channel to which the
cognitive radio is tuned. The primary base-station is assumed to be
able to track $p$ by either using a pilot signal or in a
decision-directed fashion. Thus, the cognitive radio can obtain
$\hat{p}$.
\item Upon entering the system and decoding the message of the primary user in its vicinity,
the cognitive radio simply performs amplify-and-forward relaying of
the primary codeword:\beqa X^n_c=\sqrt{\frac{P_c}{P_p}}X^n_p. \eeqa
\item The primary base-station receives
\beqa \left(p+f\sqrt{\frac{P_c}{P_p}}\right)X^n_p+Z_p^n, \eeqa hence
it can compute an estimate, $\hat{h}$, of the overall channel gain
$\left(p+f\sqrt{\frac{P_c}{P_p}}\right)$ as it decodes $m_p$.
\item The quantized version of $\hat{h}$ is then broadcast on the control channel along with
the given primary user's ID.
\item The cognitive radio picks up this information from the control channel and then computes
$\hat{h}-\hat{p}$.
\item The quantity $\hat{h}-\hat{p}$ is an estimate for $f\sqrt{P_c/P_p}$
which is then multiplied by $\sqrt{P_p/P_c}$, to obtain an estimate
for $f$.
\end{enumerate}
Note that it is possible that $\left|p+f\sqrt{P_c/P_p}\right|<|p|$
in step $3$ above. In this case the primary system would momentarily
not be able to support the requested rate of $\log(1+|p|^2P_p/N_p)$
and an Automatic Repeat Request (ARQ) would be generated by the
primary base-station. However, by this time, the cognitive radio
would have already obtained the estimate of $f$ and the next
(repeated) transmission would be guaranteed to be successful.

\appendix

\section{Proof of the converse part of
Theorem~\ref{mu_less_one}}\label{sec:proof_converse_main_result}
First we observe that the rate-region specified in
Theorem~\ref{mu_less_one} is a convex set in
Proposition~\ref{convexity}. We will use the following standard
result from convex analysis (see, for instance, \cite{Roc71}) in the
proof of the converse.

\begin{prop} A point $\mb{R}^*=(R_p^*,R_c^*)$ is on the boundary of the
a capacity region if and only if there exists a $\mu\geq 0$ such
that the linear functional $\mu R_p+R_c$ achieves its maximum, over
all $(R_p,R_c)$ in the region, at $\mb{R}^*$. \end{prop}
\subsection{The $\mu\leq 1$ case}
For convenience, we will consider a channel whose output at the
primary receiver is normalized by $a$, i.e., a channel whose
input-output single-letter equations are given by \beqa
\hat{Y}^n_p&\df& \frac{1}{a}X^n_p+X_c^n+\frac{1}{a}Z^n_p, \\
Y^n_s&=& bX^n_p+X_c^n+Z^n_s.
 \eeqa
Note that the capacity region of this channel is the same as that of
the original channel $(\ref{Y_p_standard})$, $(\ref{Y_s_standard})$
since normalization is an invertible transformation.

Suppose that a rate pair $(R_p,R_c)$ is achievable, in the sense of
Definition~\ref{ICDMS_achievability}, for the $(1,a,b,1)$-IC-DMS.
Assuming that the messages $(m_p,m_c)$ are chosen uniformly and
independently, we have, by Fano's inequality, $H(m_p|Y^n_p)\leq
n\epsilon_{p,n}$ and $H(m_c|Y^n_s)\leq n\epsilon_{s,n}$, where
$\epsilon_{p,n}\rightarrow 0$ and $\epsilon_{s,n}\rightarrow 0$ as
$\widetilde{P}^{(n)}_{e,p}\rightarrow 0$, $P^{(n)}_{e,s}\rightarrow
0$, respectively. We start with the following bound on $nR_p$: \beqa
\nonumber n R_p&\stackrel{(a)}{=}&H(m_p),\\
\nonumber &=& I(m_p;\hat{Y}^n_p) + H(m_p|\hat{Y}_p^n),\\
\nonumber &\stackrel{(b)}{\leq}& I(m_p;\hat{Y}_p^n)+n\epsilon_{p,n},\\
&=& h(\hat{Y}_p^n)-h(\hat{Y}^n_p|m_p)+n\epsilon_{p,n},
\label{Rp_converse}\eeqa where $(a)$ follows since $m_p$ and $m_c$
are uniformly distributed on $\{1,2,\ldots,2^{nR_p}\}$ and
$\{1,2,\ldots,2^{nR_p}\}$ respectively, $(b)$ follows from Fano's
inequality. Also, we have that, \beqa
\nonumber nR_c&=&H(m_c),\\
\nonumber &=& H(m_c) + H(m_c|Y^n_s,m_p)-H(m_c|Y^n_s,m_p),\\
\nonumber &=& I(m_c;Y^n_s|m_p) + H(m_c|Y^n_s,m_p),\\
\nonumber &\stackrel{(a)}{\leq}& I(m_c;Y^n_s|m_p) + n\epsilon_{s,n},\\
\nonumber &=& h(Y^n_s|m_p)-h(Y^n_s|m_p,m_c) + n\epsilon_{s,n},\\
\nonumber &\stackrel{(b)}{\leq}& h(Y^n_s|m_p)-h(Y^n_s|m_p,m_c,X^n_p,X^n_c) + n\epsilon_{s,n},\\
&\stackrel{(c)}{=}&
h(Y^n_s|m_p)-h(Z_s^n)+n\epsilon_{s,n},\label{Rc_converse} \eeqa
where $(a)$ follows from Fano's inequality and the fact that
conditioning does not increase entropy, $(b)$ follows from the fact
that conditioning does not increase entropy, and $(c)$ follows from
the the fact that $Z_s^n$ is independent of $(m_p,m_c)$ and hence
also of $(X_p^n,X_c^n)$.

Let $\widetilde{Z}^n$ be a zero mean Gaussian random vector,
independent of $(X^n_p,X^n_c,Z^n_p,Z^n_s)$ and with covariance
matrix $(\frac{1}{a^2}-1)\mb{I}_n$. Then, we can write \beqa
\nonumber h(\hat{Y}^n_p|m_p)&\stackrel{(a)}{=}& h(\hat{Y}^n_p|m_p,X^n_p),\\
\nonumber &\stackrel{(b)}{=}& h\left(\hat{Y}^n_p-\frac{1}{a}X^n_p|m_p,X^n_p\right),\\
\nonumber &=& h\left(X^n_c+\frac{1}{a}Z^n_p| m_p,X^n_p\right),\\
\nonumber &\stackrel{(c)}{=}& h(X^n_c+Z^n_s+\widetilde{Z}^n|m_p,X^n_p),\\
\nonumber &\stackrel{(d)}{=}& h(X^n_c+Z^n_s+\widetilde{Z}^n|m_p),\\
&\stackrel{(e)}{=}&
h(\widetilde{Y}^n+\widetilde{Z}^n|m_p),\label{Yp_cond} \eeqa where
$(a)$ and $(d)$ hold since $X^n_p$ is the output of a deterministic
function of $m_p$, $(b)$ holds because translation does not affect
entropy, $(c)$ follows from the fact that Gaussian distributions are
infinitely divisible and from the definition of $\widetilde{Z}^n$
and $(e)$ follows from the definition $\widetilde{Y}^n\df
X_c^n+Z_s^n$. By similar reasoning, we can write \beqa
h(Y^n_s|m_p)&=&h(\widetilde{Y}^n|m_p).\label{Ys_cond}\eeqa

Combining the bounds in $(\ref{Rp_converse})$ and
$(\ref{Rc_converse})$, we get \beqa \nonumber n(\mu R_p +R_c) &\leq&
\mu (h(\hat{Y}_p^n)-h(\hat{Y}^n_p|m_p)) +
h(Y^n_s|m_p)-h(Z_s^n)+\mu n\epsilon_{p,n}+n\epsilon_{s,n},\\
\nonumber &\stackrel{(a)}{=}& \mu h(\hat{Y}_p^n) + h(Y^n_s|m_p) -\mu
h(\hat{Y}^n_p|m_p) -
\frac{n}{2}\log(2\pi e) + \mu n\epsilon_{p,n} + n\epsilon_{s,n},\\
\nonumber &\stackrel{(b)}{=}&\mu h(\hat{Y}_p^n) +
h(\widetilde{Y}^n|m_p) -\mu h(\widetilde{Y}^n+\widetilde{Z}^n|m_p) -
\frac{n}{2}\log(2\pi e) + \mu n\epsilon_{p,n}+n\epsilon_{s,n},\\
\nonumber &\stackrel{(c)}{\leq}& \mu h(\hat{Y}_p^n)+
h(\widetilde{Y}^n|m_p)-\frac{\mu
n}{2}\log\left(e^{\frac{2}{n}h(\widetilde{Y}^n|m_p)}+
e^{\frac{2}{n}h(\widetilde{Z}^n)}\right) \\
&~&\hspace{2.5 in}-\frac{n}{2}\log(2\pi e) +\mu n\epsilon_{p,n}+
n\epsilon_{s,n},\label{lin_func_bound} \eeqa where $(a)$ follows
from the fact that $Z_s^n\sim \mathcal{N}(0,\mb{I}_{n})$, $(b)$
follows from equalities $(\ref{Yp_cond})$ and $(\ref{Ys_cond})$,
$(c)$ follows from the conditional version of the Entropy Power
Inequality (see Proposition~\ref{conditional_EPI}).

Let $X_{1}^{j-1}$ denote the first $j-1$ components of the vector
$X^n$ with the understanding that $X_1^0$ is defined to be some
constant and let $X_j$ denote the $j$-th component. We can
upper-bound $h(\widetilde{Y}^n|m_p)$ as follows: \beqa
\nonumber h(\widetilde{Y}^n|m_p) &=& h(\widetilde{Y}^n|m_p,X^n_p),\\
\nonumber &\stackrel{(a)}{=}& \sum_{j=1}^n h(\widetilde{Y}_j|m_p,\widetilde{Y}_{1}^{j-1},X_{p,j},X_{p,1}^{j-1}),\\
\nonumber &\stackrel{(b)}{\leq}& \sum_{j=1}^n h(\widetilde{Y}_j|X_{p,j}),\\
\nonumber &\stackrel{(c)}{\leq}& \sum_{j=1}^n
\frac{1}{2}\log\left(2\pi e\left(\expect[\widetilde{Y}_j^2]-
\frac{\expect[\widetilde{Y}_j
X_{p,j}]^2}{\expect[X_{p,j}^2]}\right)\right),\\
&\stackrel{(d)}{=}& \sum_{j=1}^n\frac{1}{2}\log\left(2\pi
e\left((1-\alpha_j)P_{c,j}+1\right)\right),\label{jointly_gaussian}\\
&\stackrel{(e)}{\leq}& \frac{n}{2}\log\left(2\pi
e\left((1-\alpha)P_{c}+1\right)\right), \label{Y_cond_bound}\eeqa
 where $(a)$ follows from the chain rule and $(b)$ follows from the
 fact that conditioning does not increase entropy, and $(c)$ follows
 from Lemma~\ref{prop_conditional}. Equality $(d)$ follows from
 the following argument: Since jointly Gaussian $X_{p,j}$,
$Y_{p,j}$ achieve equality in $(c)$ (by
Lemma~\ref{prop_conditional}), we can without loss of generality,
let \beqa
X_{c,j}=\hat{X}_{c,j}+\sqrt{\alpha_j\frac{P_{c,j}}{P_{p,j}}}X_{p,j},
\eeqa where $\hat{X}_{c,j}\sim\mathcal{N}(0,(1-\alpha_j) P_{c,j})$
is independent of $X_{p,j}$ and \beqa P_{c,j}\df
\frac{1}{2^{nR_c}}\sum_{j=1}^{2^{nR_c}}X_{c,j}^2,\quad P_{p,j}\df
\frac{1}{2^{nR_p}}\sum_{j=1}^{2^{nR_p}}X_{p,j}^2. \eeqa The
parameter $\alpha_j\in[0,1]$ is chosen so that the resulting
covariance $K_{X_{p,j},X_{c,j},Y_{s,j},Y_{p,j}}$ is the same as that
induced by the code. Inequality labeled with $(e)$ follows from
Jensen's inequality, by choosing $\alpha\in [0,1]$ such that \beqa
\alpha P_c=\frac{1}{n}\sum_{j=1}^n\alpha_j P_{c,j}, \eeqa and from
the fact that the power constraint $\|X^n_c\|^2\leq nP_c$ implies
that $\frac{1}{n}\sum_{j=1}^n P_{c,j}=P_c$.

Similarly, we can upper bound $h(\hat{Y}_p)$ as follows: \beqa
\nonumber h(\hat{Y}^n_p)&\stackrel{(a)}{=}& \sum_{j=1}^n
h(\hat{Y}_{p,j}|\hat{Y}_{p,1}^{j-1}),\\
\nonumber &\stackrel{(b)}{\leq}& \sum_{j=1}^n h(\hat{Y}_{p,j}),\\
\nonumber &\stackrel{(c)}{\leq}& \sum_{j=1}^n \frac{1}{2}\log(2\pi e\expect[\hat{Y}_{p,j}^2]),\\
\nonumber &\stackrel{(d)}{=}& \sum_{j=1}^n
\frac{1}{2}\log\left(\frac{2\pi e}{a^2}\left(P_{p,j}+2\sqrt{\alpha_j
P_{p,j} P_{c,j}}+
P_{c,j}+1\right)\right),\\
&\stackrel{(e)}{\leq}& \frac{n}{2}\log\left(\frac{2\pi
e}{a^2}\left((\sqrt{P_{p}}+\sqrt{\alpha
P_{c}})^2+(1-\alpha)P_{c}+1\right)\right),\label{Yp_bound}
 \eeqa
where $(a)$ follows from the chain rule and $(b)$ follows from the
 fact that conditioning does not increase entropy, $(c)$ holds since the Gaussian
 distribution maximizes the differential entropy for a fixed variance, $(d)$ follows from
 the same argument as in $(\ref{jointly_gaussian})$ and $(e)$ comes from Jensen's inequality applied
 to the $\log(\cdot)$ and the $\sqrt{\cdot}$ functions.

 Let $f(x)\df x
-\frac{\mu
n}{2}\log\left(e^{\frac{2}{n}x}+e^{\frac{2}{n}h(\widetilde{Z}^n)}\right)$
over $x\in \mathbb{R}$. Then, we can express the bound on our linear
functional in $(\ref{lin_func_bound})$ as \beqa n(\mu R_p +R_c)
&\leq& \mu h(\hat{Y}_p^n)+ f(h(\widetilde{Y}^n|m_p))
-\frac{n}{2}\log(2\pi e) + \mu n\epsilon_{p,n} + n\epsilon_{s,n}.
\label{f_bound}\eeqa Observe that as long as $\mu\leq 1$, $f(x)$ is
increasing. Hence we can obtain a further upper bound by
substituting inequalities $(\ref{Y_cond_bound})$ and
$(\ref{Yp_bound})$ into $(\ref{f_bound})$:

\beqa n(\mu R_p +R_c) &\leq& \mu \frac{n}{2}\log\left(\frac{2\pi
e}{a^2}\left((\sqrt{P_{p}}+\sqrt{\alpha
P_{c}})^2+(1-\alpha)P_{c}+1\right)\right)\\
&~&+ f\left(\frac{n}{2}\log\left(2\pi
e\left((1-\alpha)P_{c}+1\right)\right)\right) -\frac{n}{2}\log(2\pi
e) + \mu n \epsilon_{p,n} + n\epsilon_{s,n},\\
&\stackrel{(a)}{=}& \mu \frac{n}{2}\log\left(\frac{2\pi
e}{a^2}\left((\sqrt{P_{p}}+\sqrt{\alpha
P_{c}})^2+(1-\alpha)P_{c}+1\right)\right)\\
&~& + \frac{n}{2}\log\left(2\pi
e\left((1-\alpha)P_{c}+1\right)\right)-\mu\frac{n}{2}\log\left(2\pi
e\left((1-\alpha)P_c+\frac{1}{a^2}\right)\right)
\\
&~& -\frac{n}{2}\log(2\pi e) + \mu n\epsilon_{p,n} +
n\epsilon_{s,n},
 \eeqa
where $(a)$ follows from the fact that \beqa f(x)&=& x-\frac{\mu
n}{2}\log\left(e^{\frac{2}{n}x}+e^{\frac{2}{n}h(\widetilde{Z}^n)}\right),\\
&=& x-\frac{\mu n}{2}\log\left(e^{\frac{2}{n}x}+2\pi
e\left(\frac{1}{a^2}-1\right)\right), \eeqa which holds since
$\widetilde{Z}^n$ is zero mean Gaussian with covariance
$\left(\frac{1}{a^2}-1\right)\mb{I}$.

Grouping together the $\mu$-terms, dividing by $n$ and letting
$n\rightarrow \infty$, we get that \beqa \mu R_p + R_c
&\leq&\frac{\mu}{2}\log\left(1+\frac{(\sqrt{P_p}+a\sqrt{\alpha
P_c})^2}{1+a^2(1-\alpha)P_c}\right)+
\frac{1}{2}\log\left(1+(1-\alpha)P_c\right). \label{mu_sum}\eeqa Let
$\alpha_\mu$ denote the maximizing $\alpha\in [0,1]$ for a given
$\mu\leq 1$ in the above expression. Then, we can write \beqa \mu
R_p + R_c
&\leq&\frac{\mu}{2}\log\left(1+\frac{(\sqrt{P_p}+a\sqrt{\alpha_\mu
P_c})^2}{1+a^2(1-\alpha_\mu)P_c}\right)+
\frac{1}{2}\log\left(1+(1-\alpha_\mu)P_c\right).
\label{mu_sum_2}\eeqa Hence we have established the converse of the
theorem for $\mu\leq 1$.

\subsection{The $\mu>1$ case}\label{converse_mu_larger_one}
\subsubsection{Proof outline}
Suppose that ``genie A'' gives the message $m_p$ to the cognitive
receiver. We will refer to this channel as the IC-DMS(A). The
capacity region of the IC-DMS(A) must contain the capacity region of
the original IC-DMS.

\begin{prop} The capacity region of the
$(1,a,0,1)$-IC-DMS(A) is identical to the capacity region of
$(1,a,b,1)$-IC-DMS(A) for every $b\in\mathbb{R}$ and every
$a\in\mathbb{R}$.
\end{prop}
\noindent{\em Proof:} Since $m_p$ is known at the secondary receiver
along with the primary encoding rule $e^n_p$, the secondary receiver
of the $(1,a,0,1)$-IC-DMS(A) can form $bX^n_p$ and add it to its
received signal $Y^n_s$. The result is statistically identical to
the the output at the secondary receiver of the
$(1,a,b,1)$-IC-DMS(A). Thus the capacity region is independent of
$b$.\hfill $\square$

This proposition allows us to set $b=0$ without loss of generality
in any IC-DMS(A).

Now suppose that ``genie B'' gives $m_c$ to the primary transmitter
of the $(1,a,0,1)$-IC-DMS(A). We will refer to this channel as the
$(1,a,0,1)$-IC-DMS(A,B) and we note that its capacity region must
contain the capacity region of the original $(1,a,b,1)$-IC-DMS as
well as that of the IC-DMS(A). Observe that this channel is
equivalent to a broadcast channel with two antennas at the
transmitter and one antenna at each of the receivers ($2\times 1$
MIMO BC channel) with per-antenna power constraints but {\em with
additional knowledge} of $m_p$ at the secondary receiver.

\begin{figure}[h!]
\begin{center}
  \includegraphics[width=15.5cm]{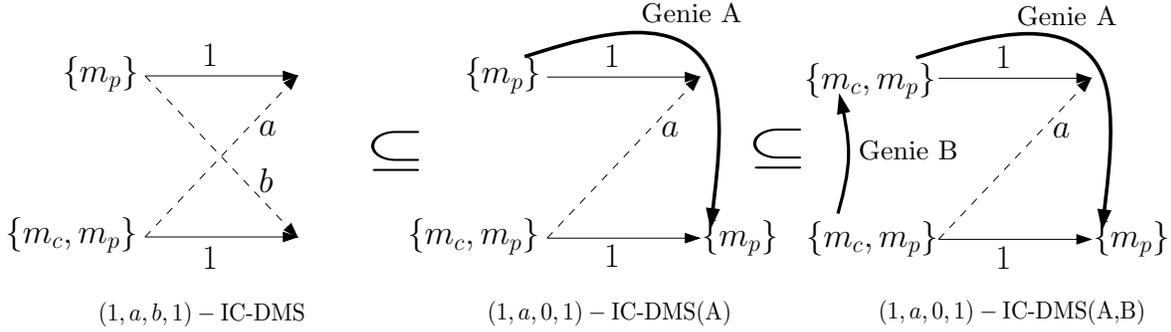}\\
  \caption{The $(1,a,b,1)$-IC-DMS, the $(1,a,0,1)$-IC-DMS(A) and the $(1,a,0,1)$-IC-DMS(A,B)
  channels and the relationships between their capacity regions.}
  \end{center}
\end{figure}

Thus, if we can show that the rates achieved by our proposed scheme
for the $(1,a,b,1)$-IC-DMS (given by $(\ref{R_p_low})$ and
$(\ref{R_c_low})$) are optimal for the $(1,a,0,1)$-IC-DMS(A,B), then
we are done. To this end, we will first define a sequence of
channels -- each of which has a capacity region that includes the
capacity region of the $(1,a,0,1)$-IC-DMS(A,B) -- such that the
rates $(\ref{R_p_low})$ and $(\ref{R_c_low})$ are optimal in the
limit.

\subsubsection{The aligned $(1,a,0,1)$-IC-DMS(A,B): The achievability}\label{IC-DMS(B)}

Consider the following modification of the $(1,a,0,1)$-IC-DMS(A,B):
Add one antenna at each of the receivers so that the input-output
relationship becomes \beqa \mb{Y}_p&=&\begin{bmatrix}
                                       1 & a \\
                                       1 & 0 \\
                                     \end{bmatrix}\mb{X}+\mb{Z}_p,\label{AMBC_Yp}\\
                             \mb{Y}_s&=&\begin{bmatrix}
                                       \epsilon & 1 \\
                                       0 & 1 \\
                                     \end{bmatrix}\mb{X}+\mb{Z}_s,\label{AMBC_Yc}\eeqa
where $\epsilon>0$ and $a\neq 0$. The vectors $\mb{Z}_p$ and
$\mb{Z}_s$ are distributed according to
$\mathcal{N}(0,\mb{\Sigma}_z)$ (their cross-correlation is
irrelevant), where \beqa
\mb{\Sigma}_z=\begin{bmatrix}1 & 0 \\
                      0 & M \\
                      \end{bmatrix},
\eeqa for some $M>0$. As in the original $(1,a,0,1)$-IC-DMS(A,B),
the message $m_p$ is known at the secondary receiver. Clearly, the
capacity region of this channel contains the capacity region of the
$(1,a,0,1)$-IC-DMS(A,B). We shall refer to this genie-aided MIMO BC
channel as the {\em aligned} $(1,a,0,1)$-IC-DMS(A,B) in what
follows.

Let $\mb{H}_p$ and $\mb{H}_s$ denote the matrices pre-multiplying
the transmit vector $\mb{X}$ in $(\ref{AMBC_Yp})$ and
$(\ref{AMBC_Yc})$, respectively. Each coordinate of the vector
$\mb{X}\in\mathbb{R}^2$ represents the symbol on each of the
antennas and the constraint on $\mb{X}$ can in general take the form
$\expect[\mb{X}\mb{X}^T]\preceq \mb{Q}$ for some positive
semi-definite covariance constraint $\mb{Q}\succeq 0$. Let the
transmitted vector (at any time-sample) be of the form \beqa
\mb{X}=X_{p1}\mb{u}_{p1}+X_{p2}\mb{u}_{p2}+X_{c1}\mb{u}_{c1}+X_{c2}\mb{u}_{c2},\label{Gaussian_strategy}
\eeqa where $\mb{u}_{p1},\mb{u}_{p2}\in\mathbb{R}^2$ and
$\mb{u}_{c1},\mb{u}_{c2}\in\mathbb{R}^2$ are the so-called signature
vectors and symbols $X_{p1},X_{p2}$ and $X_{c1},X_{c2}$ are i.i.d.
$\mathcal{N}(0,1)$.

In order to emulate the per-user individual power constraints of the
IC-DMS, we impose the per-antenna constraints
$(\expect[\mb{X}\mb{X}^T])_{11}\leq P_p$ and
$(\expect[\mb{X}\mb{X}^T])_{22}\leq P_c$ on the achievable
strategies in MIMO BC channel. We let \beqa
\mb{\Sigma}_p&\df&\mb{u}_{p1}\mb{u}_{p1}^T+\mb{u}_{p2}\mb{u}_{p2}^T,\\
\mb{\Sigma}_c&\df&\mb{u}_{c1}\mb{u}_{c1}^T+\mb{u}_{c2}\mb{u}_{c2}^T,\eeqa
so that, by the independence of $X_{p1},X_{p2},X_{c1}$ and $X_{c2}$,
the constraint can be expressed as
$(\mb{\Sigma}_p+\mb{\Sigma}_c)_{11}\leq P_p$ and
$(\mb{\Sigma}_p+\mb{\Sigma}_c)_{22}\leq P_c$.

Substituting the expression for $\mb{X}$ given in
$(\ref{Gaussian_strategy})$, the channel equations become \beqa
\mb{Y}_p&=&\mb{H}_p(X_{p1}\mb{u}_{p1}+X_{p2}\mb{u}_{p2}) +
\mb{H}_p(X_{c1}\mb{u}_{c1}+X_{c2}\mb{u}_{c2})+
\mb{Z}_p,\label{Yp_MIMO}\\
\mb{Y}_s&=&\mb{H}_s(X_{p1}\mb{u}_{p1}+X_{p2}\mb{u}_{p2}) +
\mb{H}_s(X_{c1}\mb{u}_{c1}+X_{c2}\mb{u}_{c2})+
\mb{Z}_s.\label{Yc_MIMO} \eeqa

Consider the following encoding scheme: first choose $X_{p1}$ and
$X_{p2}$ to be independent and distributed according to
$\mathcal{N}(0,1)$, and then perform Costa precoding to encode the
information in $(X_{c1},X_{c2})$ treating the interference
$\mb{H}_s(X_{p1}\mb{u}_{p1}+X_{p2}\mb{u}_{p2})$ as side-information
known at the transmitter\footnote{Costa's scheme is a block-coding
scheme and, strictly speaking, encoding is performed on the vector
$(X_{c1}^n,X_{c2}^n)$ given $X_{p1}^n$ and $X_{p2}^n$.}. The rates
achievable with such a scheme are: \beqa
R_p=R_p(\mb{\Sigma}^*_p,\mb{\Sigma}^*_c)&\df&
\frac{1}{2}\log\left|\mb{I}+(\mb{I}+
\mb{\Sigma}_z^{-1}\mb{H}_p\mb{\Sigma}^*_c\mb{H}_p^T)^{-1}\mb{\Sigma}_z^{-1}\mb{H}_p\mb{\Sigma}^*_p\mb{H}_p^T\right|
,\label{ADBC_rates_Rp}\\
R_c=R_c(\mb{\Sigma}^*_p,\mb{\Sigma}^*_c)&\df&
\frac{1}{2}\log\left|\mb{I}+\mb{\Sigma}_z^{-1}\mb{H}_s\mb{\Sigma}^*_c\mb{H}_s^T\right|,
\label{ADBC_rates_Rc} \eeqa where $\mb{\Sigma}_p^*$ and
$\mb{\Sigma}_c^*$ are the solutions of \beqa
\arg\max_{(\mb{\Sigma}_p,\mb{\Sigma}_c)\in \mathcal{S}(P_p,P_c)} \mu
R_p(\mb{\Sigma}_p,\mb{\Sigma}_c)+R_c(\mb{\Sigma}_p,\mb{\Sigma}_c),\label{opt_covariance}
\eeqa where $\mu
> 1$ and $\mathcal{S}(P_p,P_c)\df\{\mb{\Sigma}_p\succeq
0,\mb{\Sigma}_c\succeq 0: (\mb{\Sigma}_p+\mb{\Sigma}_c)_{11}\leq
P_p,~(\mb{\Sigma}_p+\mb{\Sigma}_c)_{22}\leq P_c\}$.

Since the per-antenna power constraints must be met with
equality,\footnote{If, instead, antenna $1$ uses only $P_p-\eta$
power, we can add another antenna with power $\eta$ whose signal the
receivers can first decode and then subtract off thus boosting at
least one of the rates. The same applies to antenna $2$.} we can,
without loss of generality, write \beqa
\mb{\Sigma}_p&=&\begin{bmatrix}
                \beta P_p & k_p \\
                      k_p & \alpha P_c \\
            \end{bmatrix},\quad \text{where }k_p\in
\left[-\sqrt{\alpha\beta P_p P_c},\sqrt{\alpha\beta P_p P_c}\right],\label{Sigma_p}\\
\mb{\Sigma}_c&=&\begin{bmatrix}
                        (1-\beta) P_p & k_c \\
                        k_c & (1-\alpha) P_c \\
                        \end{bmatrix},\quad
\text{where }k_c\in \left[-\sqrt{\bar{\alpha}\bar{\beta} P_p
P_c},\sqrt{\bar{\alpha}\bar{\beta} P_p P_c}\right],\label{Sigma_c}
\eeqa and $\beta\in [0,1]$, $\alpha\in [0,1]$ and $\bar{\alpha}\df
1-\alpha$, $\bar{\beta}\df 1-\beta$. With $\mb{\Sigma}_c$ expressed
in this way, we obtain \beqa \lim_{M\rightarrow
\infty}\lim_{\epsilon\rightarrow
0}\mb{\Sigma}_z^{-1}\mb{H}_s\mb{\Sigma}_c\mb{H}_s^T=
\begin{bmatrix}
                                                                 (1-\alpha)P_c & (1-\alpha)P_c \\
                                                                 0 & 0 \\
                                                               \end{bmatrix},
\eeqa in $(\ref{ADBC_rates_Rc})$. Similarly, by direct matrix
calculations we get \beqa \lim_{M\rightarrow
\infty}\lim_{\epsilon\rightarrow
0}(\mb{I}+\mb{\Sigma}_z^{-1}\mb{H}_p\mb{\Sigma}_c\mb{H}_p^T)^{-1}&=&
\begin{bmatrix}
  \frac{1}{(1-\beta)P_p+2ak_c+a^2(1-\alpha)P_c+1} &
  \frac{-(1-\beta)P_c-ak_c}{(1-\beta)P_p+2ak_c+a^2(1-\alpha)P_c+1} \\
  0 & 1 \\
\end{bmatrix},\\
\lim_{M\rightarrow \infty}\lim_{\epsilon\rightarrow
0}\mb{\Sigma}_z^{-1}\mb{H}_p\mb{\Sigma}_p\mb{H}_p^T&=&
\begin{bmatrix}
\beta P_p +2ak_p+a^2\alpha P_c & \beta P_p + ak_p\\
0 & 0
\end{bmatrix}.
\eeqa Hence, on the one hand we have, by the continuity of
$R_c(\mb{\Sigma}_p,\mb{\Sigma}_c)$ in $M$ and $\epsilon$, that \beqa
\lim_{M\rightarrow \infty}\lim_{\epsilon\rightarrow
0}R_c(\mb{\Sigma}_p,\mb{\Sigma}_c)=
\frac{1}{2}\log(1+(1-\alpha)P_c),\eeqa for any choice of $\beta\in
[0,1]$. On the other hand, we have, by the continuity of
$R_p(\mb{\Sigma}_p,\mb{\Sigma}_c)$ in $M$ and $\epsilon$, that
\beqa\lim_{M\rightarrow \infty}\lim_{\epsilon\rightarrow
0}R_p(\mb{\Sigma}_p,\mb{\Sigma}_c)=\frac{1}{2}\log\left(1+\frac{\beta
P_p+2ak_p+a^2\alpha
P_c}{(1-\beta)P_p+2ak_c+a^2(1-\alpha)P_c+1}\right).\label{limit_Rp}
\eeqa The limiting rate $(\ref{limit_Rp})$ is maximized by choosing
$\beta=1$ (and, therefore, $k_c=0$) and $k_p=\sqrt{\alpha P_p P_c}$.
Thus, \beqa \mb{\Sigma}^*_p&=&\begin{bmatrix}
                                        P_p & \sqrt{\alpha P_p P_c}  \\
                                        \sqrt{\alpha P_p P_c} & \alpha P_c \\
                                    \end{bmatrix},\\
\mb{\Sigma}^*_c&=&\begin{bmatrix}0 & 0  \\
                               0 & (1-\alpha) P_c \\
                                    \end{bmatrix},
\eeqa which is achieved by simply choosing \beqa
\mb{u}^*_{p1}=\begin{bmatrix}
                \sqrt{P_p} \\
                \sqrt{\alpha P_c} \\
              \end{bmatrix},\quad
\mb{u}^*_{c1}=\begin{bmatrix}
                0 \\
                \sqrt{(1-\alpha) P_c} \\
              \end{bmatrix},\quad \mb{u}^*_{p2}=\mb{0},\quad
              \mb{u}^*_{c2}=\mb{0}.\eeqa
Therefore, in the limit as $M\rightarrow \infty$ and
$\epsilon\rightarrow 0$, this scheme achieves the rates given by
$(\ref{R_p_low})$ and $(\ref{R_c_low})$ in the aligned
$(1,a,0,1)$-IC-DMS(A,B).

\subsubsection{The aligned $(1,a,0,1)$-IC-DMS(A,B): The converse}\label{IC-DMS(A,B)_converse} Since both
$\mb{H}_p$ and $\mb{H}_s$ are invertible for every $\epsilon>0$ and
$a\neq 0$, we can equivalently represent this channel by the
equations \beqa
\widetilde{\mb{Y}}_p&=&\mb{X}+\widetilde{\mb{Z}}_p,\label{AMBC_Yp2}\\
\widetilde{\mb{Y}}_s&=&\mb{X}+\widetilde{\mb{Z}}_s.\label{AMBC_Yc2}\eeqa
The new noise vectors are given by
$\widetilde{\mb{Z}}_p\sim\mathcal{N}(0,\mb{H}_p^{-1}\mb{\Sigma}_z\mb{H}_p^{-T})$
and
$\widetilde{\mb{Z}}_s\sim\mathcal{N}(0,\mb{H}_s^{-1}\mb{\Sigma}_z\mb{H}_s^{-T})$.
This channel is then exactly in the form of an Aligned MIMO BC
channel (AMBC) (see \cite{WSS04}, Section~2), but with $m_p$
revealed to the secondary receiver.

Let $\widetilde{\mb{Y}}^n_p\in \mathbb{R}^{2\times n}$ and
$\widetilde{\mb{Y}}^n_s\in\mathbb{R}^{2\times n}$ denote the channel
outputs over a block of $n$ channel uses. We can upper bound any
achievable rate $R_p$ as follows \beqa
nR_p&=&H(m_p),\\
&=& I(m_p;\widetilde{\mb{Y}}^n_p) + H(m_p|\widetilde{\mb{Y}}_p^n),\\
&\stackrel{(a)}{\leq}&
I(m_p;\widetilde{\mb{Y}}_p^n)+n\widetilde{\epsilon}_{p,n}, \eeqa
where $(a)$ follows from Fano's inequality with
$\widetilde{\epsilon}_{p,n}\rightarrow 0$ as $n\rightarrow \infty$.
Noting that the secondary receiver observes the tuple
$(\widetilde{\mb{Y}}^n_s,m_p)$, we can write \beqa
nR_c&=&H(m_c),\\
&=& H(m_c) + H(m_c|(\widetilde{\mb{Y}}^n_s,m_p))-H(m_c|(\widetilde{\mb{Y}}^n_s,m_p)),\\
&=& I(m_c;(\widetilde{\mb{Y}}^n_s,m_p)) + H(m_c|(\widetilde{\mb{Y}}^n_s,m_p)),\\
&\stackrel{(a)}{\leq}& I(m_c;(\widetilde{\mb{Y}}^n_s,m_p)) +
n\widetilde{\epsilon}_{s,n},\\
&\stackrel{(b)}{=}& I(m_c;\widetilde{\mb{Y}}^n_s|m_p) +
n\widetilde{\epsilon}_{s,n}, \eeqa where $(a)$ follows from Fano's
inequality with $\widetilde{\epsilon}_{s,n}\rightarrow 0$ as
$n\rightarrow \infty$, and $(b)$ follows since $I(m_p;m_c)=0$.

Thus, we can upper-bound the linear functional of the achievable
rates as \beqa \mu R_p +R_c&\leq& \nonumber
\frac{\mu}{n}I(m_p;\widetilde{\mb{Y}}^n_p)+\frac{1}{n}I(m_c;\widetilde{\mb{Y}}^n_s|m_p)+
\mu\widetilde{\epsilon}_{p,n}+\widetilde{\epsilon}_{s,n},\\
&=& \frac{\mu}{n} h(\widetilde{\mb{Y}}^n_p)-\frac{\mu}{n}
h(\widetilde{\mb{Y}}^n_p|m_p)+\frac{1}{n}h(\widetilde{\mb{Y}}^n_s|m_p)
-\frac{1}{n}h(\widetilde{\mb{Z}}^n_s)+
\mu\widetilde{\epsilon}_{p,n}+\widetilde{\epsilon}_{s,n}\label{ADBC_converse}
\eeqa where $\mu >1$.

Now, from Proposition~4.2 of \cite{WSS04}, we know that, for every
$\mu>1$, there exists an {\em enhanced} Aligned Degraded BC channel
(ADBC) which contains the capacity region of the AMBC given by
$(\ref{AMBC_Yp2})$ and $(\ref{AMBC_Yc2})$, and for which the maximum
of the linear functional $\mu R_p +R_c$, over all $(R_p,R_c)$ in the
region, is equal to the maximum of the same linear functional over
the capacity region of the corresponding AMBC (i.e., the two regions
meet at the point of tangency). Due to the degradedness, we can
write the channel outputs of the enhanced ADBC as \beqa
\bar{\mb{Y}}^n_s&=&\mb{X}^n+\bar{\mb{Z}}^n_s,\\
\bar{\mb{Y}}^n_p&=&\bar{\mb{Y}}^n_s+\bar{\mb{Z}}_p^n,\\
 \eeqa
where the matrices $\bar{\mb{Z}}^n_s$ and $\bar{\mb{Z}}^n_p$ are
constructed such that their columns, denoted by $\bar{\mb{Z}}_s$ and
$\bar{\mb{Z}}_p$, are independent, zero-mean Gaussian with
covariances satisfying $\mb{\Sigma}_{\bar{\mb{Z}}_s}\preceq
\mb{\Sigma}_{\widetilde{\mb{Z}}_s}$ and
$\mb{\Sigma}_{\bar{\mb{Z}}_s}+\mb{\Sigma}_{\bar{\mb{Z}}_p}\preceq
\mb{\Sigma}_{\widetilde{\mb{Z}}_p}$ (see proof of Proposition~4.2 of
\cite{WSS04} for how to construct them). Hence, for this enhanced
ADBC, we can write $(\ref{ADBC_converse})$ as \beqa \nonumber \mu
R_p +R_c&\leq& \frac{\mu}{n}
h(\bar{\mb{Y}}^n_p)+\frac{1}{n}h(\bar{\mb{Y}}^n_s|m_p)-\frac{\mu}{n}
h(\bar{\mb{Y}}^n_p|m_p) -\frac{1}{n}h(\bar{\mb{Z}}^n_s)+
\mu\bar{\epsilon}_{n}\\
\nonumber &=& \frac{\mu}{n}
h(\bar{\mb{Y}}^n_s+\bar{\mb{Z}}^n_p)+\frac{1}{n}h(\bar{\mb{Y}}^n_s|m_p)-\frac{\mu}{n}
h(\bar{\mb{Y}}^n_s+\bar{\mb{Z}}^n_p|m_p)
-\frac{1}{n}h(\bar{\mb{Z}}^n_s)+
\mu\bar{\epsilon}_{n},\\
\nonumber &\leq& \frac{\mu}{n}
h(\bar{\mb{Y}}^n_p)+\frac{1}{n}h(\bar{\mb{Y}}^n_s|m_p)
-\mu\log\left(e^{\frac{2}{2n}h(\bar{\mb{Y}}^n_s|m_p)}+
e^{\frac{2}{2n}h(\bar{\mb{Z}}^n_p)}\right)\\
&~&\hspace{3.4in}-\frac{1}{n}h(\bar{\mb{Z}}^n_s)+
\mu\bar{\epsilon}_{n},\label{enhanced_converse}
 \eeqa
where we have used the conditional version of the vector Entropy
Power Inequality (see Proposition~\ref{prop_conditional}) in the
last step.

The key property of the this enhanced ADBC is that the upper bound
$(\ref{enhanced_converse})$ is maximized by choosing the input
$\mb{X}$ to be Gaussian, i.e., the vector EPI is tight (see proof of
Theorem~3.1 of \cite{WSS04}). Hence, an optimal achievable scheme
for this ADBC is the Costa precoding strategy\footnote{Note that for
the ADBC a simple superposition scheme is also optimal.} that is
described in Section~\ref{IC-DMS(B)}: The largest jointly achievable
rates are given by \beqa R_p&=&R_p(\mb{\Sigma}^*_p,\mb{\Sigma}^*_c),\\
R_c&=&R_c(\mb{\Sigma}^*_p,\mb{\Sigma}^*_c)\eeqa where
$R_p(\mb{\Sigma}^*_p,\mb{\Sigma}^*_c)$ and
$R_c(\mb{\Sigma}^*_p,\mb{\Sigma}^*_c)$ are as given by
$(\ref{ADBC_rates_Rp})$ and $(\ref{ADBC_rates_Rc})$, respectively.

Since this scheme is also achievable for the AMBC, the capacity
region of the ADBC and AMBC are identical (see Theorem~4.1 of
\cite{WSS04}). Moreover, it is obvious that this scheme is also
achievable for the AMBC with {\em additional knowledge} of $m_p$ at
the secondary receiver: The knowledge of $m_p$ is simply ignored by
the receiver. Hence, this scheme is optimal for the aligned
$(1,a,0,1)$-IC-DMS(A,B) (as defined by $(\ref{AMBC_Yp})$ and
$(\ref{AMBC_Yc})$) with $\mu>1$ as well.

Since the Pareto-optimal (for $\mu>1$) rates for the limiting (as
$M\rightarrow \infty$ and $\epsilon\rightarrow 0$) aligned
$(1,a,0,1)$-IC-DMS(A,B) exactly match the rates $(\ref{R_p_low})$
and $(\ref{R_c_low})$ achievable in the original $(1,a,b,1)$-IC-DMS
channel, {\em and} since the capacity region of the
$(1,a,b,1)$-IC-DMS is contained in the capacity region of the
aligned $(1,a,0,1)$-IC-DMS(A,B) for any $M,\epsilon>0$, we have
completed the proof of the converse part of
Theorem~\ref{mu_less_one} for $\mu>1$.

\section{Proof of Corollary~\ref{cor_sum}}\label{proof_sum_capacity}

The proof of this Corollary follows from Theorem~\ref{mu_less_one}
and Lemma~\ref{lemma_sum_cap}. In particular, we observe that the
converse to Theorem~\ref{mu_less_one} for $\mu\geq 1$ (see
Section~\ref{converse_mu_larger_one}) holds for any $a>0$ and
$b\in\mathbb{R}$. However, from Lemma~\ref{lemma_sum_cap} we see
that the choice $\alpha=1$ in $(\ref{R_p_low})$ and
$(\ref{R_c_low})$ is optimal for any $a\geq 1$, as long as $\mu\geq
1$. Hence the corollary is proved.\hfill $\square$.

\noindent {\bf Remark:} This result implies that, for any $a\geq 1$,
$b\in\mathbb{R}$ and $\mu\geq 1$, the linear functional $\mu
R_p+R_c$ is maximized at $(R_p,R_c)=(C_{\text{sum}}(a),0)$. Hence,
for $a\geq 1$, the entire capacity region is parametrized by
$\mu\leq 1$, for any $b\in\mathbb{R}$.

\section{Proof of the converse part of
Theorem~\ref{thm:high_interference}}\label{sec:converse_high_interference}
Let ``genie B'' disclose $m_c$ to the primary transmitter, thus
getting a $2\times 1$ MIMO BC channel with per-antenna power
constraints. The input-output relationship for this channel can be
written as \beqa Y_p&=&\mb{h}_p^T\mb{X}+Z_p,\\
Y_s&=&\mb{h}_s^T\mb{X}+Z_s,
 \eeqa
where $\mb{h}_p=[1~~a]^T$ and $\mb{h}_s=[b~~1]^T$. We choose
$\mu\leq 1$ in the linear functional $\mu R_p+R_c$ and recall that
the optimal transmission vector $\mb{X}$ is Gaussian and given by
$(\ref{Gaussian_strategy})$ and the optimal encoding strategy is to
generate $X_p$ by Costa precoding for
$\mb{h}_p^T(X_{c1}\mb{u}_{c1}+X_{c2}\mb{u}_{c2})$ (see
\cite{WSS04}). Consequently, in place of $(\ref{ADBC_rates_Rp})$ and
$(\ref{ADBC_rates_Rc})$, we get, respectively, \beqa R_p=
\hat{R}_p(\mb{\Sigma}^*_p,\mb{\Sigma}^*_c)&\df&\frac{1}{2}\log\left(1+\mb{h}_p^T\mb{\Sigma}^*_p\mb{h}_p\right),\label{MIMO_Rp2}\\
R_c= \hat{R}_c(\mb{\Sigma}^*_p,\mb{\Sigma}^*_c)&\df&
\frac{1}{2}\log\left(1+\frac{\mb{h}_s^T\mb{\Sigma}^*_c\mb{h}_s}
{1+\mb{h}_s^T\mb{\Sigma}^*_p\mb{h}_s}\right)\label{MIMO_Rc2}, \eeqa
where $\mb{\Sigma}_c^*$ and $\mb{\Sigma}_c^*$ are the solutions of
$(\ref{opt_covariance})$ but with $\mu \leq 1$. Substituting the
covariance matrices $(\ref{Sigma_p})$ and $(\ref{Sigma_c})$ into
$(\ref{MIMO_Rp2})$ and $(\ref{MIMO_Rc2})$, we get \beqa
\hat{R}_p(\mb{\Sigma}_p,\mb{\Sigma}_c)=\hat{R}_p(\beta,\alpha,k_p,a,b)&\df&\frac{1}{2}\log\left(1+\beta
P_p +2ak_p+\alpha a^2P_c\right),\label{Rp_high}\\
\hat{R}_c(\mb{\Sigma}_p,\mb{\Sigma}_c)=\hat{R}_c(\beta,\alpha,k_p,a,b)&\df&\frac{1}{2}\log\left(1+\frac{b^2(1-\beta)P_p+2k_cb+(1-\alpha)P_c}
{1+b^2\beta P_p +2k_pb+\alpha P_c}\right).\label{Rc_high} \eeqa The
expression in $(\ref{Rc_high})$ is maximized by choosing
$k_c=\sqrt{(1-\beta)(1-\alpha)P_pP_c}$, i.e., making $\mb{\Sigma}_c$
unit rank. If $b=0$ it is clear that $\beta=1$ and $k_p=\sqrt{\alpha
P_p P_c}$ maximizes the linear functional $\mu
\hat{R}_p(\beta,\alpha,k_p,a,b)+\hat{R}_c(\beta,\alpha,k_p,a,b)$. In
general, we would like to find the set of all values of $b$ for
which $\beta=1$ and $k_p=\sqrt{\alpha P_p P_c}$ are optimal. For
such values of $b$, we then have \beqa
\hat{R}_p(\mb{\Sigma}_p,\mb{\Sigma}_c)&=&\frac{1}{2}\log\left(1+\left(\sqrt{P_p}+a\sqrt{\alpha P_c}\right)^2\right),\\
\hat{R}_c(\mb{\Sigma}_p,\mb{\Sigma}_c)&=&\frac{1}{2}\log\left(1+\frac{(1-\alpha)P_c}
{1+\left(b\sqrt{P_p} +\sqrt{\alpha P_c}\right)^2}\right), \eeqa
which exactly match the achievable rates given in
Lemma~\ref{lemma_achievable_high}. To this end, let $B(\mu,a)$
denote the set of all $b>0$ such that the function \beqa
\max_{0\leq\alpha\leq 1}\mu
\hat{R}_p\left(\beta,\alpha,k_p,a,b\right)
+\hat{R}_c(\beta,\alpha,k_p,a,b)\label{B_function}\eeqa is
maximized, over all $\beta\in [0,1]$ and $k_p\in [-\sqrt{\beta\alpha
P_p P_c},\sqrt{\beta\alpha P_p P_c}]$, by choosing $\beta=1$ and
$k_p=\sqrt{\alpha P_p P_c}$. We let $b_{\max }(\mu,a)\df
\max_{b\in B(\mu,a)}$ to obtain the statement of the theorem. %Numerical evaluation of $b_{\max }$ indicates
%that the maximization of the function $(\ref{B_function})$ can be
%carried out only over $\beta$ by fixing $k_p=\sqrt{\alpha P_p P_c}$,
%i.e., we can fix $\mb{\Sigma}_p$ to be unit rank without loss of
%generality.
Appealing to the remark in the proof of Corollary~\ref{cor_sum} (see
Appendix~\ref{proof_sum_capacity}), we observe that the boundary of
the capacity region in this very-high-interference-gain regime is
completely parametrized by $\mu\leq 1$. Hence, we have proved the
theorem.

\section{Supporting results}\label{sec:support}
\begin{prop}\label{convexity}
The rate region specified in Theorem~\ref{mu_less_one} is a convex
set.
\end{prop}
\noindent{\em Proof:} A point $\mb{R}=(R_p,R_c)$ is in the rate
region specified in Theorem~\ref{mu_less_one} if and only if there
exists $\alpha\in [0,1]$ such that \beqa 0\leq R_c&\leq&
\frac{1}{2}\log(1+(1-\alpha)P_c),\label{R_c_temp}\\
0\leq R_p &\leq& \frac{1}{2}\log\left(1+a^2P_c+P_p+2a\sqrt{\alpha
P_pP_c}\right)+\frac{1}{2}\log\left(\frac{1}{1+a^2(1-\alpha)P_c}\right).\label{R_sum_temp}
 \eeqa
Suppose that there exist two points
$\mb{R}^{(1)}=(R_p^{(1)},R_c^{(1)})$ and
$\mb{R}^{(2)}=(R^{(2)}_p,R^{(2)}_c)$ that are in the region. Let
$\alpha^{(1)}\in [0,1]$ and $\alpha^{(2)}\in [0,1]$ be their
corresponding parameters in $(\ref{R_c_temp})$ and
$(\ref{R_sum_temp})$. Then for any $\lambda\in [0,1]$, we have that
\beqa \lambda R_c^{(1)}+(1-\lambda)R_c^{(2)}&\leq&
\frac{\lambda}{2}\log(1+(1-\alpha^{(1)})P_c)+\frac{1-\lambda}{2}\log(1+(1-\alpha^{(2)})P_c),\\
&\leq&\frac{1}{2}\log(1+(1-\alpha^*)P_c) \eeqa where $\alpha^*\df
\lambda\alpha^{(1)}+(1-\lambda)\alpha^{(2)}$ and the last inequality
follows from Jensen's inequality. Similarly, \beqa \lambda
R^{(1)}_p+(1-\lambda)R^{(2)}_p &\leq&
\left[\frac{\lambda}{2}\log\left(1+a^2P_c+P_p+2a\sqrt{\alpha^{(1)}
P_pP_c}\right)\right.\\
&~&\hspace{0.3in}\left.+\frac{1-\lambda}{2}\log\left(1+a^2P_c+P_p+2a\sqrt{\alpha^{(2)}
P_p P_c}\right)\right]\\
&~&\hspace{0.3in}+\left[\frac{\lambda}{2}\log\left(\frac{1}{1+a^2(1-\alpha^{(1)})P_c}\right)\right.\\
&~&\hspace{1.8in}\left.+\frac{1-\lambda}{2}\log\left(\frac{1}{1+a^2(1-\alpha^{(2)})P_c}\right)\right],\\
&\stackrel{(a)}{\leq}& \frac{1}{2}\log\left(1+a^2P_c+P_p+2a\sqrt{P_p
P_c}\left(\lambda\sqrt{\alpha^{(1)}}+(1-\lambda)\sqrt{\alpha^{(2)}}\right)\right)\\
&~&\hspace{1in}+\frac{1}{2}\log\left(\frac{1}{1+a^2(1-\lambda
\alpha^{(1)}-(1-\lambda)\alpha^{(2)})P_c}\right),\\
&\stackrel{(b)}{\leq}& \frac{1}{2}\log\left(1+a^2P_c+P_p+2a\sqrt{P_p
P_c\alpha^*}\right)+\frac{1}{2}\log\left(\frac{1}{1+a^2(1-\alpha^*)P_c}\right).
\eeqa $(a)$ follows from Jensen's inequality applied to the concave
function $\log(k_1+k_2x)$ (for constant $k_1,k_2>0$) and the concave
function $\log\left(\frac{1}{1+(1-x)k}\right)$ (for constant $k>0$).
Inequality $(b)$ follows from Jensen's inequality applied to the
square-root function. Hence $\lambda
\mb{R}^{(1)}+(1-\lambda)\mb{R}^{(2)}$ is in the region as well,
hence the region is a convex set.\hfill$\square$

\begin{prop}[Conditional EPI]\label{conditional_EPI}
Suppose $Y^n\in\mathbb{R}^n$ and $Z^n\in\mathbb{R}^n$ are
independent random vectors and $m\in\{1,2,\ldots,M\}$ (for some $M$)
is independent of $Z^n$. Then we have that \beqa h(Y^n+Z^n|m)\geq
\frac{n}{2}\log\left(e^{\frac{2}{n}h(Y^n|m)}+e^{\frac{2}{n}h(Z^n)}\right).
\eeqa
\end{prop}
\noindent{\em Proof:} \beqa h(Y^n+Z^n|m)&=&\sum_{i=1}^M
h(Y^n+Z^n|m=i)\prob(m=i),\\
&\stackrel{(a)}{\geq}&\sum_{i=1}^M
\frac{n}{2}\log\left(e^{\frac{2}{n}h(Y^n|m=i)}+e^{\frac{2}{n}h(Z^n)}\right)\prob(m=i),\\
&\stackrel{(b)}{\geq}&\frac{n}{2}\log\left(e^{\frac{2}{n}h(Y^n|m)}+e^{\frac{2}{n}h(Z^n)}\right),
\eeqa where $(a)$ follows from the classical Entropy Power
Inequality (EPI) (see e.g. \cite{Dembo}), and $(b)$ follows from
Jensen's inequality applied to the convex function
$\log(e^{2x/n}+k)$ (for constant $k$ and $n$).\hfill $\square$.

\begin{lemma}\label{prop_conditional}
Given two zero-mean random variables $X$ and $Y$ with a fixed
covariance matrix $K_{XY}$ we have that \beqa h(Y|X)\leq
\frac{1}{2}\log\left(2\pi e
\left(\expect[Y^2]-\frac{\expect[YX]^2}{\expect[X^2]}\right)\right),
\eeqa with equality when $X$ and $Y$ are jointly Gaussian.
\end{lemma}
\noindent{\em Proof:} Let $\beta=\frac{\expect[XY]}{\expect[X^2]}$.
Then the MMSE estimator of $Y$ given $X$ is given by $\hat{Y}=\beta
X$. \beqa
h(Y|X) &\stackrel{(a)}{=}& h(Y-\beta X|X),\\
&\stackrel{(b)}{\leq}& h(Y-\beta X),\\
&\stackrel{(c)}{\leq}& \frac{1}{2}\log\left(2\pi e\left(\expect[(Y-\beta X)^2]\right)\right),\\
&=& \frac{1}{2}\log\left(2\pi
e\left(\expect[Y^2]-\frac{\expect[XY]^2}{\expect[X^2]}\right)\right),
\eeqa where $(a)$ follows from the fact that shifts do not change
the differential entropy, $(b)$ follows since conditioning does not
increase entropy, and $(c)$ follows since the Gaussian distribution
maximizes the entropy for a given variance. By the orthogonality
principle, $(b)$ is tight when $X$ and $Y$ are jointly Gaussian and
in that case $(c)$ is tight as well.\hfill $\square$

\begin{lemma}\label{lemma_sum_cap}
\beqa &~&\hspace{-0.8in}\max_{0\leq\alpha\leq 1}
\frac{\mu}{2}\log\left(1+\frac{(\sqrt{P_p}+a\sqrt{\alpha
P_c})^2}{1+a^2(1-\alpha)P_c}\right)+
\frac{1}{2}\log\left(1+(1-\alpha)P_c\right)\label{maximization}\\
\nonumber &~&\hspace{2.5in}=
\frac{\mu}{2}\log\left(1+\left(\sqrt{P_p}+a\sqrt{P_c}\right)^2\right),
\eeqa for $a\geq 1$ and $\mu\geq 1$.
\end{lemma}
\noindent{\em Proof:} On the one hand we have that \beqa
&~&\max_{0\leq\alpha\leq 1}
\frac{\mu}{2}\log\left(1+\frac{(\sqrt{P_p}+a\sqrt{\alpha
P_c})^2}{1+a^2(1-\alpha)P_c}\right)+
\frac{1}{2}\log\left(1+(1-\alpha)P_c\right),\\
&~&\hspace{0.8in}= \max_{0\leq \alpha\leq
1}\frac{1}{2}\log\left(\frac{\left(1+a^2(1-\alpha)P_c+(\sqrt{P_p}+a\sqrt{\alpha
P_c})^2\right)^\mu(1+(1-\alpha)P_c)}{(1+a^2(1-\alpha)P_c)^\mu}\right),\\
&~&\hspace{0.8in}\leq \max_{0\leq \alpha\leq
1}\frac{1}{2}\log\left(\frac{\left(1+a^2(1-\alpha)P_c+(\sqrt{P_p}+a\sqrt{\alpha
P_c})^2\right)^\mu}{(1+a^2(1-\alpha)P_c)^{\mu-1}}\right),\\
&~&\hspace{0.8in}= \max_{0\leq \alpha\leq
1}\frac{1}{2}\log\left(\frac{\left(1+a^2P_c+P_p+2a\sqrt{\alpha
P_pP_c}\right)^\mu}{(1+a^2(1-\alpha)P_c)^{\mu-1}}\right),\\
&~&\hspace{0.8in}=\frac{\mu}{2}\log\left(1+\left(\sqrt{P_p}+a\sqrt{P_c}\right)^2\right).
\eeqa On the other hand, the maximization problem in
$(\ref{maximization})$ can be lower bounded with
$\frac{\mu}{2}\log\left(1+\left(\sqrt{P_p}+a\sqrt{P_c}\right)^2\right)$,
by choosing $\alpha=1$. Hence the lemma is proved.\hfill $\square$

\section*{Acknowledgement}
The authors would like to thank Patrick Mitran for his comments on
the previous draft of this paper.

\end{document}